\documentclass[amsmath, 
               amssymb, 
               prx,
               floatfix,
               preprint,
               superscriptaddress]{revtex4-2}
\usepackage{graphicx}
\usepackage{xcolor}
\usepackage{ulem}
\usepackage{mathrsfs}

\usepackage{microtype}
\usepackage{placeins}
\usepackage{array}
\usepackage{tabularx}
\usepackage{booktabs}

\newcommand{\avg}[1]{\left\langle #1 \right\rangle}
\newcommand{\ens}[1]{\left\lbrace #1 \right\rbrace}

\DeclareTextFontCommand{\emph}{\itshape}

\begin{document}

\title{Mapping Inter-City Trade Networks to Maximum Entropy Models
       using Electronic Invoice Data
    }

\author{Cesar I. N. Sampaio Filho}
\affiliation{
    Departamento de Física, Universidade Federal do Ceará, 60451-970
    Fortaleza, Ceará, Brazil
}
\affiliation{
    Centro de Análise de Dados e Avaliação de Políticas Públicas,
    Instituto de Pesquisa e Estratégia Econômica do Ceará, 60822-325,
    Fortaleza, Ceará, Brazil
}
\author{Rilder S. Pires}
\affiliation{
    Laboratório de Ciência de Dados e Inteligência
    Artificial, Universidade de Fortaleza, 60811-905 Fortaleza, Ceará, Brazil 
}
\affiliation{
    Centro de Análise de Dados e Avaliação de Políticas Públicas,
    Instituto de Pesquisa e Estratégia Econômica do Ceará, 60822-325,
    Fortaleza, Ceará, Brazil    
}

\author{Humberto A. Carmona} 
\affiliation{
    Departamento de Física, Universidade Federal do Ceará, 60451-970
    Fortaleza, Ceará, Brazil    
}
\affiliation{
    Centro de Análise de Dados e Avaliação de Políticas Públicas,
    Instituto de Pesquisa e Estratégia Econômica do Ceará, 60822-325,
    Fortaleza, Ceará, Brazil    
}
\author{José S. Andrade Jr.}
\email[]{soares@fisica.ufc.br}
\affiliation{
    Departamento de Física, Universidade Federal do Ceará, 60451-970
    Fortaleza, Ceará, Brazil    
}
\affiliation{
    Centro de Análise de Dados e Avaliação de Políticas Públicas,
    Instituto de Pesquisa e Estratégia Econômica do Ceará, 60822-325,
    Fortaleza, Ceará, Brazil    
}

\date{\today}

\begin{abstract}
We analyze the network of transactions among cities based on the electronic
invoice database for the municipalities in the Ceará state, Brazil. This
database consists of approximately \(3.7\) billion records, each containing 43
fields of information, registered during the period between the years 2016
to 2019.
All the transactions are grouped in a unique dataset and represented as an
asymmetrical adjacency matrix corresponding to a directed graph with
connections weighted by the number of transactions among cities. Due to the
large size of Ceará state, \(148,894.442~km^{2}\)~\cite{ce_area}, its
unequal distribution of wealth, and spatially heterogeneous population
density, we initially determine communities of cities based on their mutual
intensity of trades and then verify to which extent their economic interests
somehow reflect what we define as a “community cohesiveness”.
For the first task, we use the Infomap algorithm to detect the partition which
provides the shortest description length and captures the optimal community
structure of the network in terms of its associated flow dynamics.
Surprisingly, the partition identified has five modules, whose
two-dimensional geographical projections are all simply-connected domains,
\textit{i.e.}, consisting of single pieces without holes.
Having described the topological properties of the transaction network, we
proceed with the analysis of our database from the perspective of traded
products by building bipartite structures represented in terms of adjacency
matrices between municipalities and products, considering both the contexts
of selling and buying.
We then make use of the revealed comparative advantage (RCA) concept, widely
used in foreign trade analyses, to define a non-monetary and binary activity
index that is capable to distinguish the relative advantage of a city in a
class of goods or services as evidenced by trade flows.
Finally, through the pairwise Maximum Entropy Model, we can associate to
the largest communities previously characterized, their corresponding binary
Ising-like Hamiltonian models. The local fields and couplings computed for a
given community are those that best reproduce the average product
activities of its cities as well as the statistical correlations between the
product activities of all pairs of its cities. In an analogy with critical
phenomena, our results reveal that each community operates at a
“temperature” that is close to the corresponding “critical point”,
suggesting a high degree of “economic cohesiveness” in its trade network of
cities.
\end{abstract}

\maketitle

\section{Introduction}\label{sec:introduction}

Economic geography investigates the way different regions and countries are
interconnected through trade and investment, as well as how these
relationships affect growth, development and inequality.
Understanding the complex relationships between the enormous number of
economic activities, people, firms and places requires the definition of
metrics necessarily involving dimensionality reduction techniques that
together are referred to as Economic Complexity (EC).
The Economic Complexity Index (ECI)~\cite{Hidalgo2009, Hidalgo2021} and the
Economic Fitness Index (EFI)~\cite{Tacchella2012,Tacchella2013} are two
examples of these metrics.
Both of them are based on the examination of the intrinsic interdependencies
between countries and regions by exploring international trade data. In this
way, researchers have gained insight, for example, into which countries are
important hubs in global trade network, how products can be compared in
terms of their relative complexity and distinctiveness.
Through the concept of \textit{proximity} between products, Hidalgo and
Haussmann~\cite{Hidalgo2007} proposed that the export basket of developing
countries should expand more efficiently, from an economical point of view,
to new products that are ``close'' to the ones already being exported.

One key concept present in all these approaches is that the rarity of the
products a country exports, their technological complexity and diversity
should be closely related to the country’s installed infrastructure, such as
transportation network, energy systems and intellectual
capital~\cite{Hidalgo2009, Tacchella2012}.
Using the index Revealed Comparative Advantage (RCA)~\cite{Balassa1965} to
construct a bipartite network of countries and products, it is possible to
quantify the complexity and diversity of products, so that countries exports
baskets and their Gross Domestic Product (GDP) per capita can be empirically
related, aiming to forecast countries growth~\cite{Hidalgo2009,
    Tacchella2012, Tacchella2013}.
However, it is important to note that there are many other factors
influencing a country's GDP per capita, including, for example, its natural
resources, political stability, extreme events and more.
In this way, while the export basket indeed represents a crucial factor impacting
economic growth, it is certainly not the only one.

In Ref.~\cite{Balland2022} the authors review how machine learning techniques
apply to economics with the goal of understanding the systemic interactions that
influence various socioeconomic outcomes, and discuss how big data and machine
learning are instrumental in this emerging field of economic complexity.
Brummitt et al.~\cite{Brummitt2020} introduced a machine learning technique, which
they named Principal Smooth-Dynamics Analysis (PriSDA), to explore the dynamics of
economic growth. Their findings emphasized that product diversity, particularly in
more sophisticated products, serves as a significant driver of income growth.
In a parallel vein, Albora et al.~\cite{Albora2023} employed supervised learning
techniques on the UN-COMTRADE database, using the Harmonized System 1992
classification (HS)~\cite{ASAKURA1993, wco2017}. Their research highlights that
the ability to forecast the introduction of new products is essential for
effective economic planning.
In Ref.~\cite{Jean2016}, the authors introduce a novel approach by applying
convolucional neural networks to high-resolution satellite imagery. This method
is used to estimate economic livelihood indicators, such as consumption
expenditure and asset wealth, in five developing African countries: Nigeria,
Tanzania, Uganda, Malawi, and Rwanda.

Although EC metrics were initially proposed based on international trade data, they
were further developed and applied to non-export data sets and subnacional
entities. Operti et al.~\cite{Operti2018} introduced a novel algorithm they termed
Exogenous Fitness, as an evolved form of the previously established Fitness
metric~\cite{Cristelli2013}. Focused on Brazilian states, they assessed regional
competitiveness through the export basket. By combining Exogenous Fitness scores
with GDP per capita, the authors distinguish between two economic regimes among
these states: one with high predictability and another with low predictability. The
study also compares Exogenous Fitness rankings to those from Endogenous Fitness and
the Economic Complexity Index, offering a comprehensive view of regional economic
dynamics.

In this work we apply concepts of EC in the scale of municipalities using
electronic invoice trade data.
To make an analogy with international trade, we treat sales from one
municipality to others as exports.
One relevant consideration that naturally arises is that, at this scale, it
is not guaranteed that the traded products are in fact produced locally. As
a consequence, the spatial correlations at the state level between exports
and installed capabilities is expectedly weaker, which led us to approach
the trade network from a partitioned point of view, \textit{i.e.}, by
considering the potential formation of communities.
Community structure plays a pivotal role in understanding the intricate dynamics
of complex networks, encompassing diverse domains such as social and biological
networks. Its significance lies in the substantial implications it holds for the
propagation of information, distribution of resources, and spreading of
influence within the network~\cite{Girvan2002, Bathelt2004, Peixoto2014}. In
this study, we initially investigate the phenomenon of community formation
induced by the exchange of products between municipalities in the trade network.
By examining this interplay, we aim to shed light on the underlying mechanisms
by which the ``flow'' of products induces the formation of communities.

Further, in order to understand the economic patterns commercial activities at the
scale of cities, we infer ``pairwise interactions'' between pair of cities through
the products they trade.
Particularly, using the binary municipality-product matrices of the
communities detected from the network of transactions, we find that often
the product baskets of different cities are strongly correlated, both for
sales and purchases.
We therefore exploit these statistical correlations and the average
activities of the cities using the Maximum Entropy Model (MEM) developed in
information theory. This method provides a conceptual framework based on
statistical physics models for representing a given natural process in terms
of  ``interactions'' between its elementary units using experimental
data~\cite{bialek2012biophysics}.
More precisely, the principle of maximum entropy encapsulates the core
concept behind the Inverse Ising Problem solution, or the so-called
Boltzmann machine, wherein an underlying ``Hamiltonian'' associated with a
given complex system can be inferred from the observed statistical
correlations among its constituent parts. In this way, MEM has been applied
to systems that can be mapped to Ising-like models, that is, models in which
the interacting elements are in an active or inactive state, {\it i.e.}, a
network of moments of dipole with states of spins that are up or down under
the action of an external field and their mutual interactions. In a neuronal
network, for example, interactions between pairs of neurons that react to
some stimuli are deduced from their firing
patterns~\cite{Cocco2009,Tkacik2009,Shlens2006,tang2008maximum,Mora2005,
Lotfi2020,Ioffe}. MEM have also been successful in the characterization of
protein-protein interactions~\cite{Morcos2011,Weigt2009} and genetic
interaction networks from gene expression
patterns~\cite{Stein2015,Lezon2006,Locasale2009}. Other complex systems have
been analyzed in terms of the Boltzmann machine, for example the collective
responses exhibited by flocks of  birds~\cite{Bialek2012,Bialek2014} and,
more recently, the emergence of collective behavior from the eye movement
patterns of a group of people while watching commercial
videos~\cite{Burleson2017} or reading texts~\cite{Debora2021}. 

In Ref.~\cite{Bury2013}, Bury conducts an analysis of stock market
data utilizing a maximum entropy model. Within this framework, market
indices are conceptualized as time-dependent binary spin states, each characterized
by either bullish or bearish behavior. The author focuses on two distinct financial
systems: one that encompasses eight European indices, and another that consists
of a selection of stocks from the Dow Jones index. Employing criteria outlined in
Ref.~\cite{Mora2011}, the presence of criticality in these finite
systems is investigated. Specifically, a system is deemed to be approaching a critical state if a
peak is observed in the temperature-dependent variance of the likelihood, also
known as the heat capacity, near its operational point. The study concludes that
neither system functions in a strictly critical state. The European indices
generally operate in proximity criticality, except during market
downturns. In contrast, the Dow Jones system is found to function significantly far
from criticality.

This paper is organized as follows. In Section II, we present the network of
transactions between municipalities in the Ceará state, in the Northeast of
Brazil. In Section III, we employ a community detection scheme based on the
flow dynamics in the network of transactions to identify regions in the
state that are organized by the strength of the local trade quantified by
the number of commercial transactions between different municipalities. In
Section IV, we  apply the MEM to investigate the internal ``cohesiveness''
of the largest communities focusing on the Ising-like models deduced from
the product baskets for sales and purchases of their corresponding
cities.  Finally, in Section V we present the general conclusions of the work.

\section{Networks of Transactions}\label{sec:results}

The Ceará state, in northeaster Brazil, has 184 municipalities, with an
estimated population of 9,240,580 people. The network of transactions
between pair of cities in Ceará is built from a database of all electronic
invoices~\cite{nfe2023} registered in the state between the years \(2016\) to
\(2019\), which contains approximately \(3.7\) billion records, each one
corresponding to a traded product, with \(43\) fields of information.
Precisely, we consider all the transactions of products that the cities in
Ceará carried out among themselves and with the other cities of Brazil, thus
disregarding all transactions circumscribed to a given city. We then applied
a thorough process of standardizing and sanitizing the data and deflated all
monetary values starting from January 2016, corresponding to our database's
first month. Furthermore, we only considered transactions with values larger
than \(2000\) USD\@. Once these preprocessing steps are completed, we group all
the transactions in a unique dataset and build an asymmetrical adjacency
matrix corresponding to a directed graph with connections weighted by the
number of transactions between pair of cities.

At this point, with the purpose of characterizing the economic dynamics
of the Ceará state in the time window from \(2016\) to \(2019\), we proceed with
the definition of a complex network based on the data of only internal sales
and purchase operations among cities. The network corresponds to a unique,
densely connected component with \(C=185\) nodes representing the
municipalities, and \(14479\) directed-weighted edges. The direction of the
edge is from the city that sells to the city that buys one or more given
products. To each directed edge between a pair of cities \(i\) and \(j\), a
weight is associated which corresponds to the total number of sells
(buys), \(n_{ij}\), or buys (sells), \(n_{ji}\), operations from
\(i\,(j)\) to \(j\,(i)\) performed during the time window. From this network, we
can then compute the numbers of internal selling and internal buying
connections of each city \(i\), \(k^{S}_{i}\) and \(k^{B}_{i}\), respectively, as
well as the sum of their corresponding weights,
\(W_{i}^{S}=\sum_{j=1}^{k_{i}^{S}} n_{ij}\) and
\(W_{i}^{B}=\sum_{j=1}^{k_{i}^{B}} n_{ji}\).

\begin{figure}[htpb]
    \begin{minipage}{0.47\textwidth}
        \centering
        \includegraphics*[width=\columnwidth]{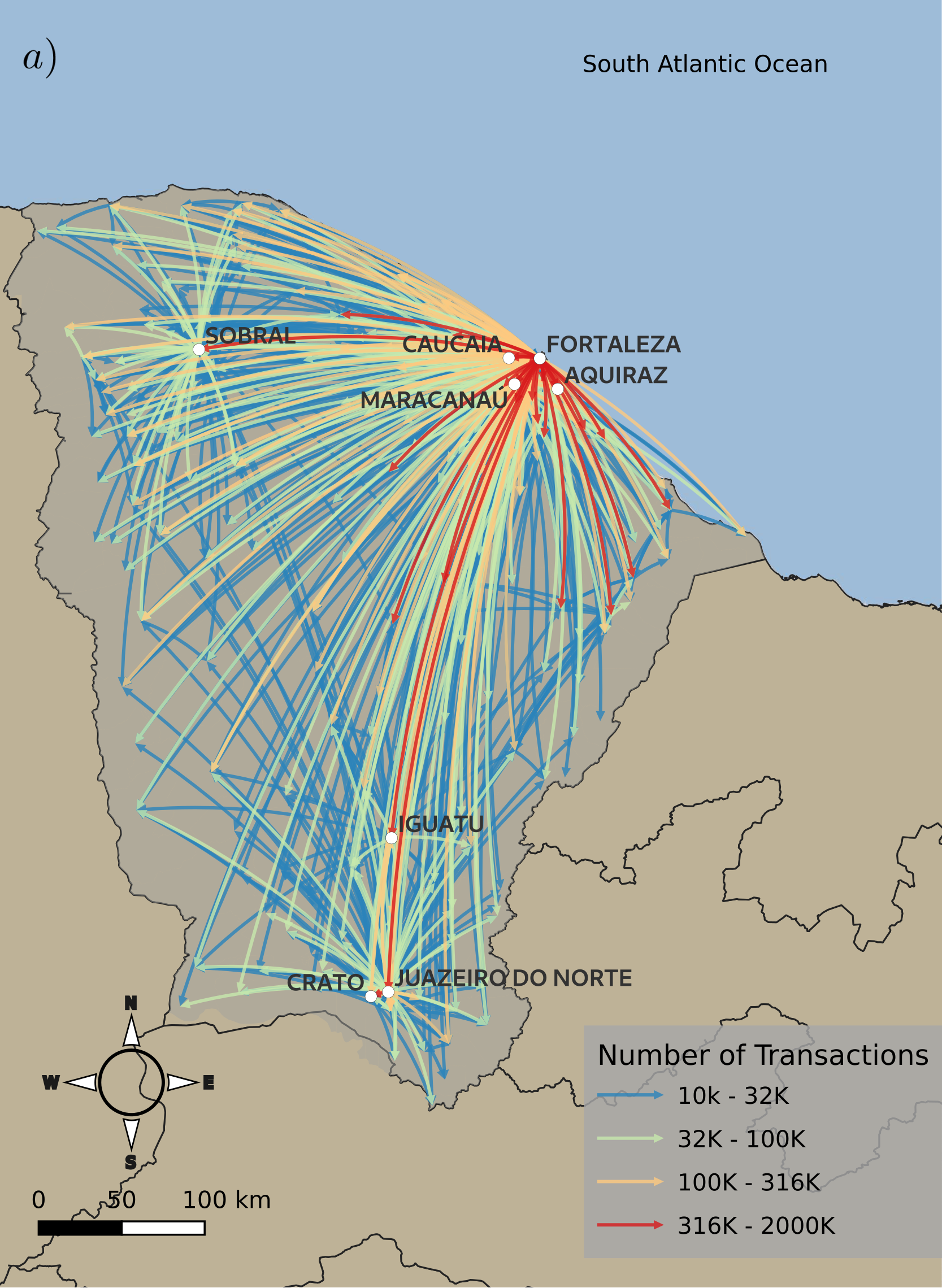}
    \end{minipage}
    \begin{minipage}{0.49\textwidth}
        \centering
        \includegraphics*[width=\columnwidth]{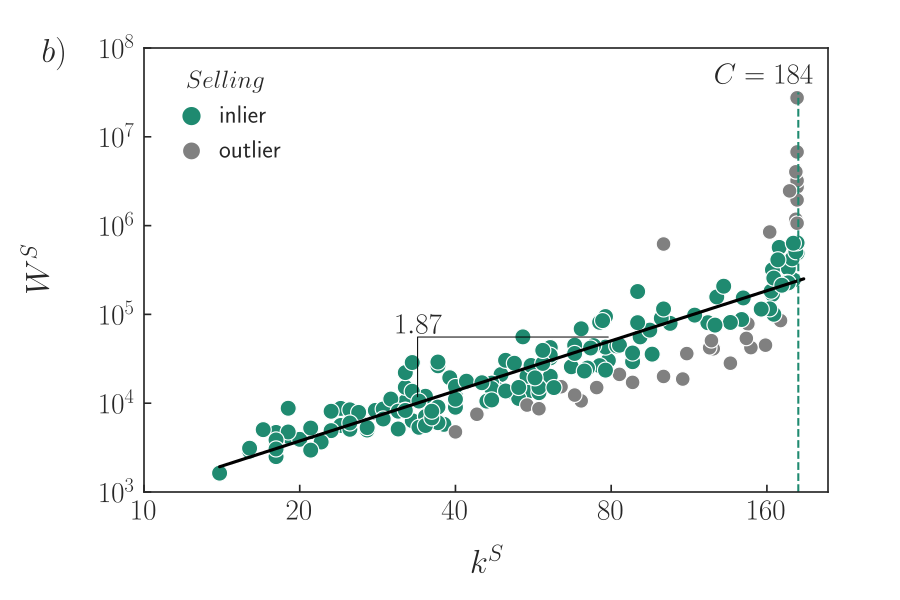}
        \includegraphics*[width=\columnwidth]{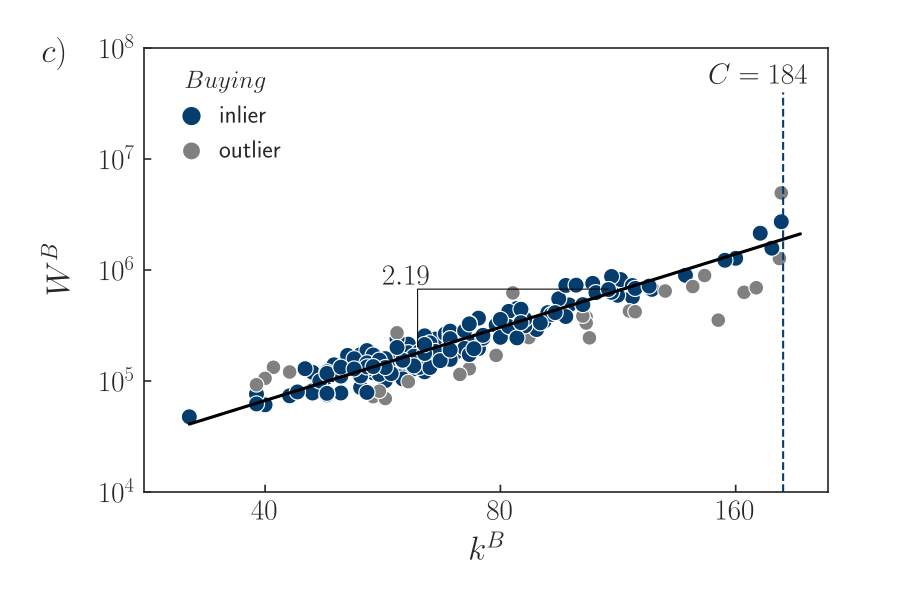}
    \end{minipage}
    \caption{a) The trade network among municipalities superimposed on the
        map of the Ceará state, Brazil. The nodes are located at the
        geographic position of each city. The color of each edge is coded
        according to the number of transactions between a pair of nodes,
        from blue to red, corresponding to low and high values,
        respectively, and its direction is from the city that sells to the
        city that buys the products. Only edges with more than \(10000\)
        transactions are displayed.\ b)~Scaling relation between the total
        number of internal sales transactions (out-transactions), \(W^{S}\),
        and the total number of internal sales connections (out-degree),
        \(k^{S}\), both computed for all cities in the state of Ceará from a
        database comprising approximately \(3.7\) billion records of
        electronic invoices registered during the period from 2016 to 2019.
        Green and gray circles correspond to the inlier and outlier data
        points, respectively, and the continuous green line represents the
        best fit to a power-law, \(W^{S}\simeq {(k^{S})}^{\beta}\), with the
        exponent \(\beta=1.87\pm 0.03\), \(95\%\) confidence interval and 
        \(R^{2} = 0.89\).\ c) The same as in (b), but for the total number of internal
        buying transactions (in-transactions), \(W^{B}\), and the total number
        of internal buying connections (in-degree), \(k^{B}\). Blue and gray
        circles correspond to the inlier and outlier data points,
        respectively, and the continuous blue line represents the best fit
        to a power-law, \(W^{B}\simeq {(k^{B})}^{\beta}\), with the exponent
        \(\beta=2.19\pm 0.01\), \(95\%\) confidence interval and 
        \(R^{2} = 0.94\).}\label{fig:fig01}
\end{figure}

Figures~\ref{fig:fig01}b and~\ref{fig:fig01}c show the dependence of the
weights of the nodes on their degrees in a double logarithmic plot. As
depicted, the relationship between these two variables seems to be well
described in terms of a power-law model for both the selling and the buying
data sets. 
In order to attenuate the effect of dispersion of the data points on the
parameter estimation, we utilized the RANSAC algorithm~\cite{Fischler1981,
Chum2008}, which statistically identifies the outliers and fits the model
considering only the inlier points. Therefore, among the data points obtained
for all cities relating the total number of their internal sales transactions
(out-transactions), \(W^{S}\), with the total number of their internal sales
connections (out-degree), \(k^{S}\), we select only the inliers (green circles)
to perform a least-squares fit to the power law, \(W^{S} \simeq
{(k^{S})}^{\beta}\), and find the exponent \(\beta=1.87\pm 0.03\). Following the
same procedure for the inliers in the plot of the in-transactions, \(W^{B}\),
against the in-degree, \(k^{B}\), the least-squares fit to the power-law,
\(W^{B}\simeq {(k^{B})}^{\beta}\), gives the exponent \(\beta=2.19\pm 0.01\).
The values of these exponents imply that the weights of the cities' connections
grow disproportionately faster than their corresponding degrees. Such a
superlinear scaling behavior contrasts with the linear one, \(W(k) \sim k\),
expected for the case in which the weights of the edges \(n\) are statistically
uncorrelated with the degree of the nodes from where they depart (selling) or to
where they arrive (buying)~\cite{Newman2004, Barrat2004, Rubinov2010,
Boccaletti2014}. 
In order to illustrate this condition, we performed additional calculations
preserving the degrees of each node \(i\), \(k_{i}^{S}\) and \(k_{i}^{B}\), but
shuffling the values of the weights \(n_{ij}\) between randomly chosen pairs of
edges in the network. In this way, strong correlations, if present in the original
network, should disappear. Indeed, the results shown in Fig.~\ref{fig:figS0} of the
supplemental material indicate that the effect of suppressing strong correlations
is to recover a linear relation between the total node weights and their degrees.

\begin{figure}[htpb]
    \begin{minipage}{0.47\textwidth}
        \centering
        \includegraphics*[width=\columnwidth]{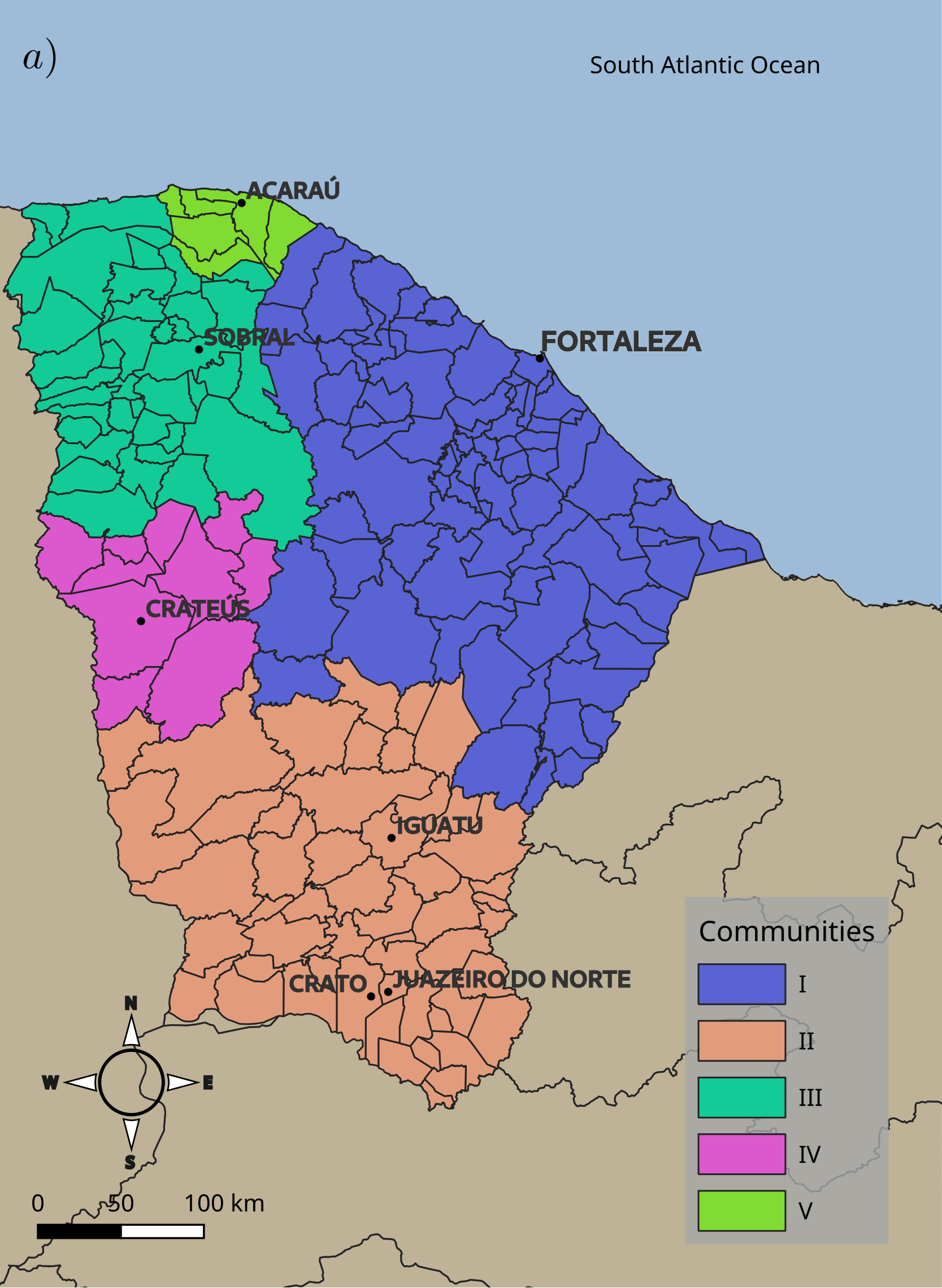}
    \end{minipage}
    \begin{minipage}{0.47\textwidth}
        \centering
        \includegraphics*[width=\columnwidth]{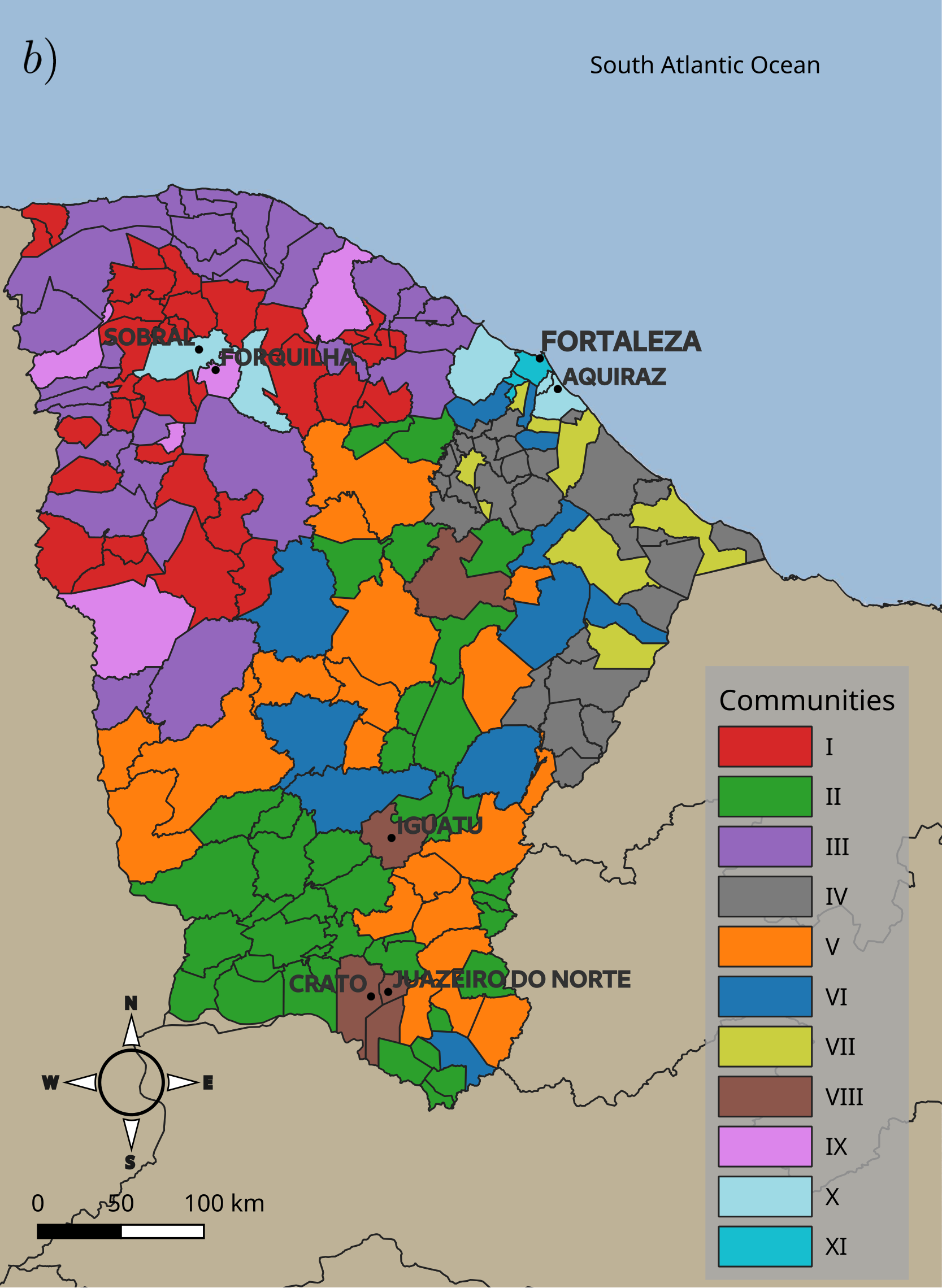}
    \end{minipage}
    \caption{Shown in (a) is the map of the Ceará state in Brazil colored
        according to the communities detected via the Infomap
        algorithm~\cite{Rosvall2008}, which identifies the partition \(M\) that
        provides the shortest description length and captures the optimal
        community structure of the network of commercial transactions between
        cities  in terms of its associated flow dynamics. As a noteworthy
        result, the algorithm is capable to identify a partition with five
        modules, whose two-dimensional geographical projections are all
        simply-connected domains, \textit{i.e.}, consisting of single pieces
        without holes. Highlighted are the leading seven cities in each
        community in terms of population and monetary resources. The results
        shown in (a) points to the reliability and validates the effectiveness
        of the flow-based Infomap algorithm~\cite{Rosvall2008}, which strongly
        contrasts with alternative methods for modularity detection basically
        relying on pairwise interactions and features of the network formation
        process, as it is the case with generalized modularity techniques
        \cite{Newman2004}. For comparison, the communities identified using the
        stochastic block model~\cite{Peixoto2014} are shown in (b).}\label{fig:fig02}
\end{figure}

Given Ceará state's vast area, alongside its uneven wealth distribution and
diverse population density patterns, our approach is twofold. First, we aim to
identify clusters of cities by examining the intensity of their mutual trades.
Subsequently, we will assess how closely their economic interests align with
what we term as community ``cohesiveness''. For the initial task, we have chosen
the Infomap algorithm,~\cite{Rosvall2008,Rosvall2011, Alzahrani2016} since it is
compatible with the most relevant and inherent feature of the system under
investigation here, namely, a network of flows, in particular, flows of
financial resources based on trade operations among cities. Accordingly, this
flow-based algorithm makes use of an information-theoretic method to detect
network communities with a very high computational performance.
Figure~\ref{fig:fig02}a shows the map of the Ceará state in Brazil colored
according to the communities detected via the Infomap algorithm. The algorithm
identifies a partition with the minimal average per-step description length of
the random walk. Interestingly, the two-dimensional geographical projections of
the resulting five modules clearly show that they are all
\textit{simply-connected domains},\textit{i.e.}, well-delimited single pieces
without holes~\cite{Nasser2020, Papamichael1981}. This result demonstrates the
consistency and corroborates the adequacy of the flow-based algorithm, which is
in evident contrast with other approaches relying on pairwise interactions and
the network formation process, as it is the case, for example, with the
generalized modularity~\cite{Fortunato2016}. For a comparison,  we show in with
Fig.~\ref{fig:fig02}b the communities detected using the stochastic block
model~\cite{Peixoto2014}. Clearly, the large number of communities, their
heterogeneity in space, and lack of contiguity among their constituting pieces
indicate that the method is unable to capture the economic dynamics embedded in
the trade network of sales and buying among cities.

\begin{figure}[htpb]
    \begin{minipage}{0.54\textwidth}
        \centering
        \includegraphics*[width=\columnwidth]{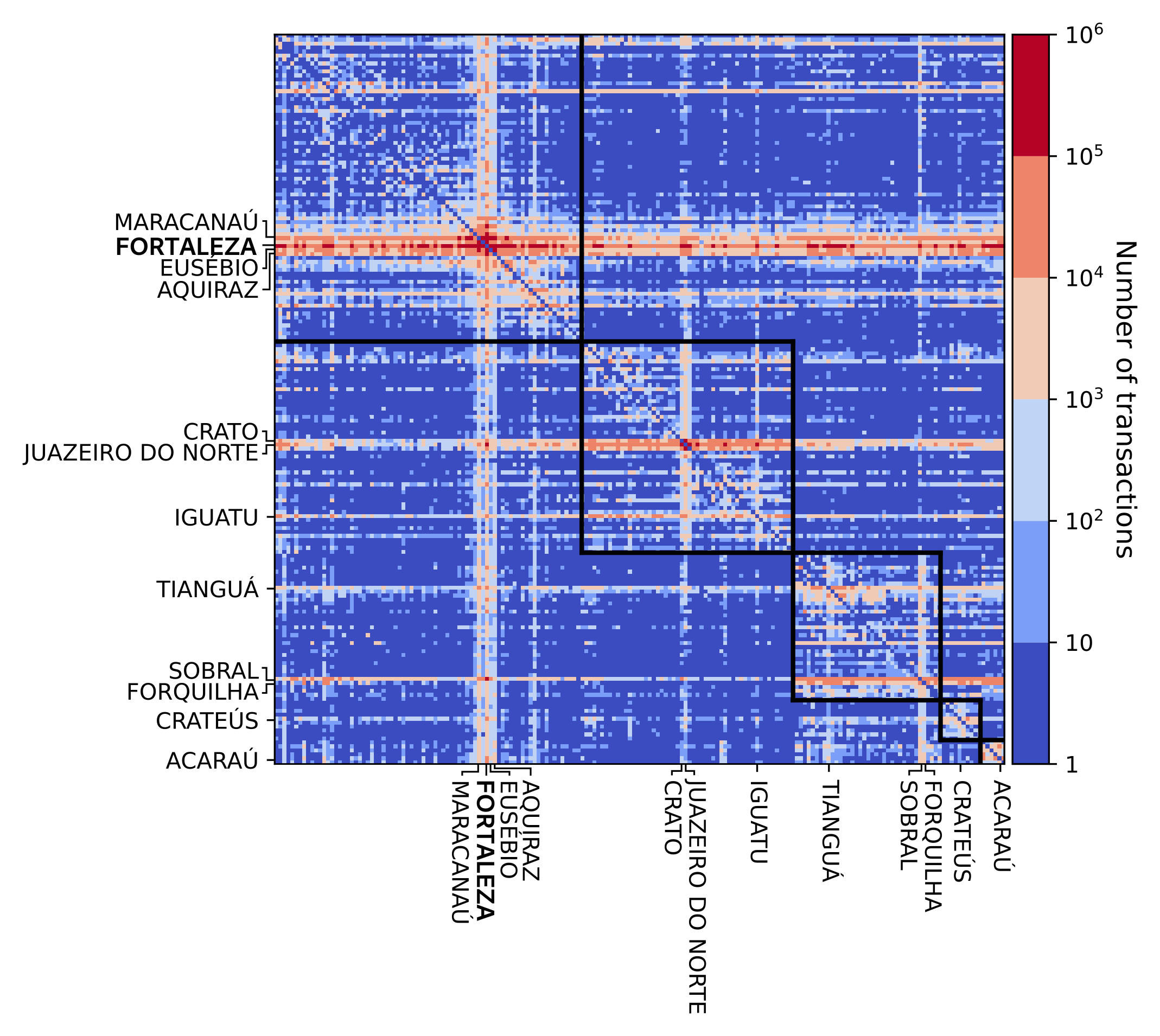}
    \end{minipage}
    \caption{The trade share matrix after
        clustering between municipalities. We sorted the cities according to
        the optimal partition found using the Infomap algorithm and
        highlighted the leading cities in terms of their number of
        transactions. The communities are separated by the solid black
        lines. The matrix has entries color-coded according to the number of
        transactions between a pair of cities, from blue to red,
        corresponding to low and high values, respectively. It is important
        to note the far-reaching strong
        connections of the highlighted cities.}\label{fig:fig03}
\end{figure}

In Fig.~\ref{fig:fig03} we show the trade share matrix after clustering
the municipalities using the Infomap, with the communities delimited by
black continuous lines. The matrix has color-coded entries according to the
number of transactions between a pair of nodes, from blue to red,
corresponding to low and high values, respectively. Moreover, it is
non-symmetrical, with rows reaching the cities that sell to the ones buying
in the columns. In each community, we highlighted the leading cities in
terms of population and monetary resources. These cities, including
Fortaleza, the capital of Ceará, present a wide spectrum transactions that
virtually extends to all other cities in the state.

\section{Community detection}

Having studied  the topological properties of the networks of transactions,
we next analyze the database of electronic invoices from the perspective of
the traded products. To this end, we build bipartite networks, one for
selling and another for buying, in which the two types of nodes are the
cities of Ceará state and the products they trade, sell or buy,
respectively, with any other city in the country. 
Following~\cite{Albora2023}, the products are identified in the electronic
invoices using the HS~\cite{ASAKURA1993, wco2017}.  
The presence of a link in these networks should reflect the relative importance
of the selling or buying transactions of a given city \(c\) involving the
product \(p\) in a balanced context of all transactions among all cities and all
products traded in the system. This is achieved here by adopting the concept of
Revealed Comparative Advantage (RCA) index [6], quantified in terms of the
complexity of products and the diversification of the cities' baskets of traded
products. More precisely, in order to determine the relevant transactions, we
adopt the following model of activation of products: let \(q_{c,p}^{S}\) be the
monetary value associated with a city \(c\) \textit{selling} a product \(p\) to
a different municipality in the state of Ceará or any other city in Brazil. Thus
\(T_{c}^{S}=\sum_{p}^{P} q_{c,p}^{S}\) corresponds to the total sales of the
city \(c\), the summation running over all \(P\) products traded in the state.
On the other hand, the amount \(T_{p}^{S} = \sum_{c}^{C}q_{c,p}^{S}\) adds up to
the total sales of a product \(p\), the summation now running over all \(C\)
municipalities in the state of Ceará. Likewise, the quantity
\(T_{tot}^{S}=\sum_{c,p}^{C,P} q_{c,p}^{S}\) amounts to the total monetary value
associated with the sales of all products \(P\) by all cities \(C\) in the state
during the investigated time window. From these three quantities, we can compute
the quotient of quotients,
\begin{equation}
    Q_{c,p}^{S}=\frac{q_{c,p}^{S}/T_{c}^{S}}{T_{p}^{S}/T_{tot}^{S}},
    \label{eq:eq01}
\end{equation}
where the numerator corresponds to the relative value of the sales of
product \(p\) by city \(c\), compared to all sales of this city. The denominator
is the relative of value of all sales of the product \(p\) to all sales of the
state.\ Equation~(\ref{eq:eq01}) is the usual definition of an element of the
RCA matrix~\cite{Balassa1965, Bernard2007, Hidalgo2007} which we are here
extending to the trades among municipalities. If \(Q_{c,p}^{S} > 1\), it means
that product \(p\) is relatively more important to the sales of city \(c\) then
the relative importance of all sales of this product by all cities. In this
case, we say that the municipality \(c\) is a ``seller'' of product \(p\).
We apply the same reasoning for \(Q_{c,p}^{B}\) representing the monetary
value associated with the purchases of product \(p\) by the municipality \(c\).
In this case, if \(Q_{c,p}^{B} > 1\), it means that buying product \(p\) is
relatively more important to the city \(c\) than the relative importance of
all purchases of this product by all cities of the state, so that the
municipality~\(c\) is a ``buyer'' of  product~\(p\). 

\begin{figure}[htpb]
    \includegraphics*[width=0.65\columnwidth]{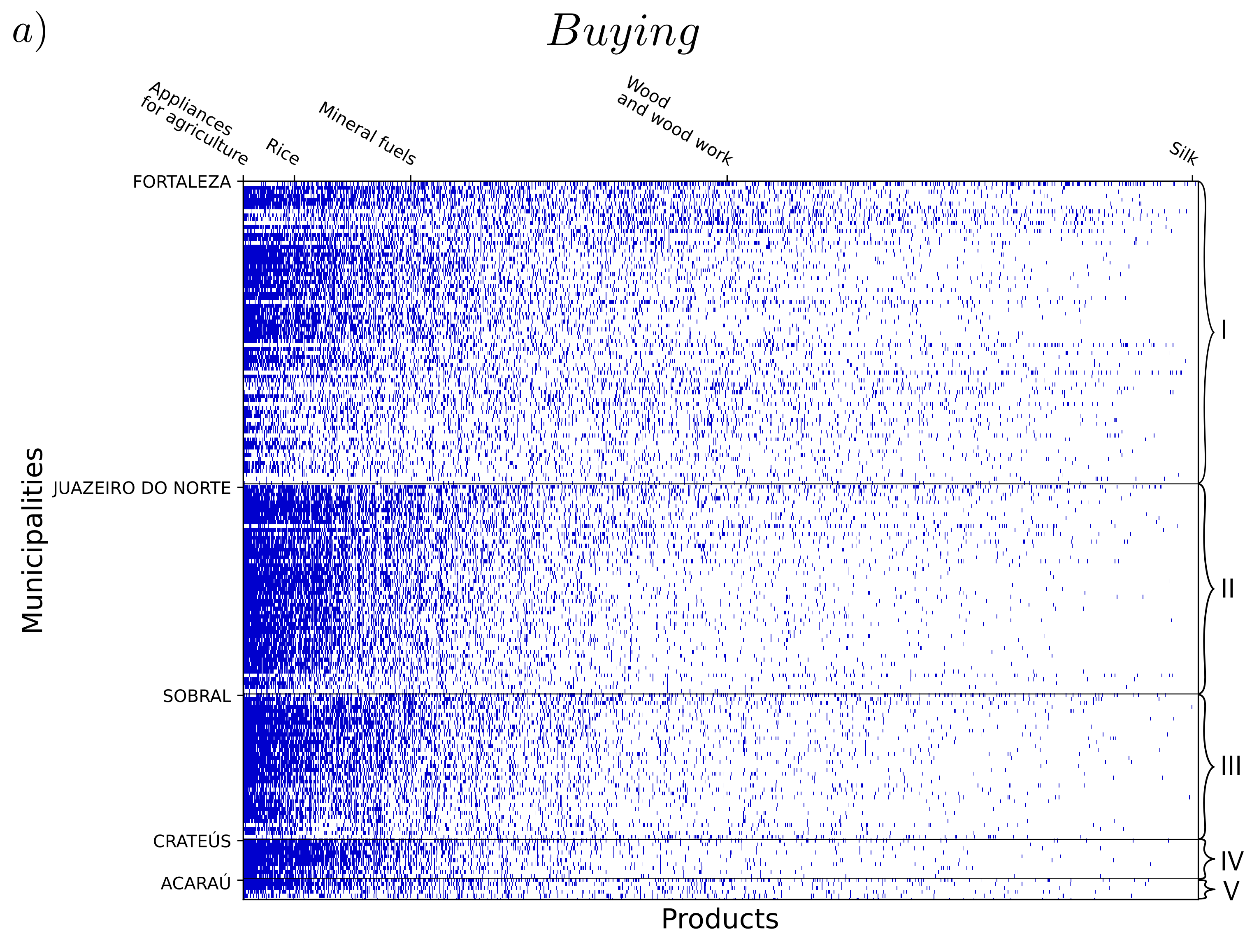}
    \includegraphics*[width=0.65\columnwidth]{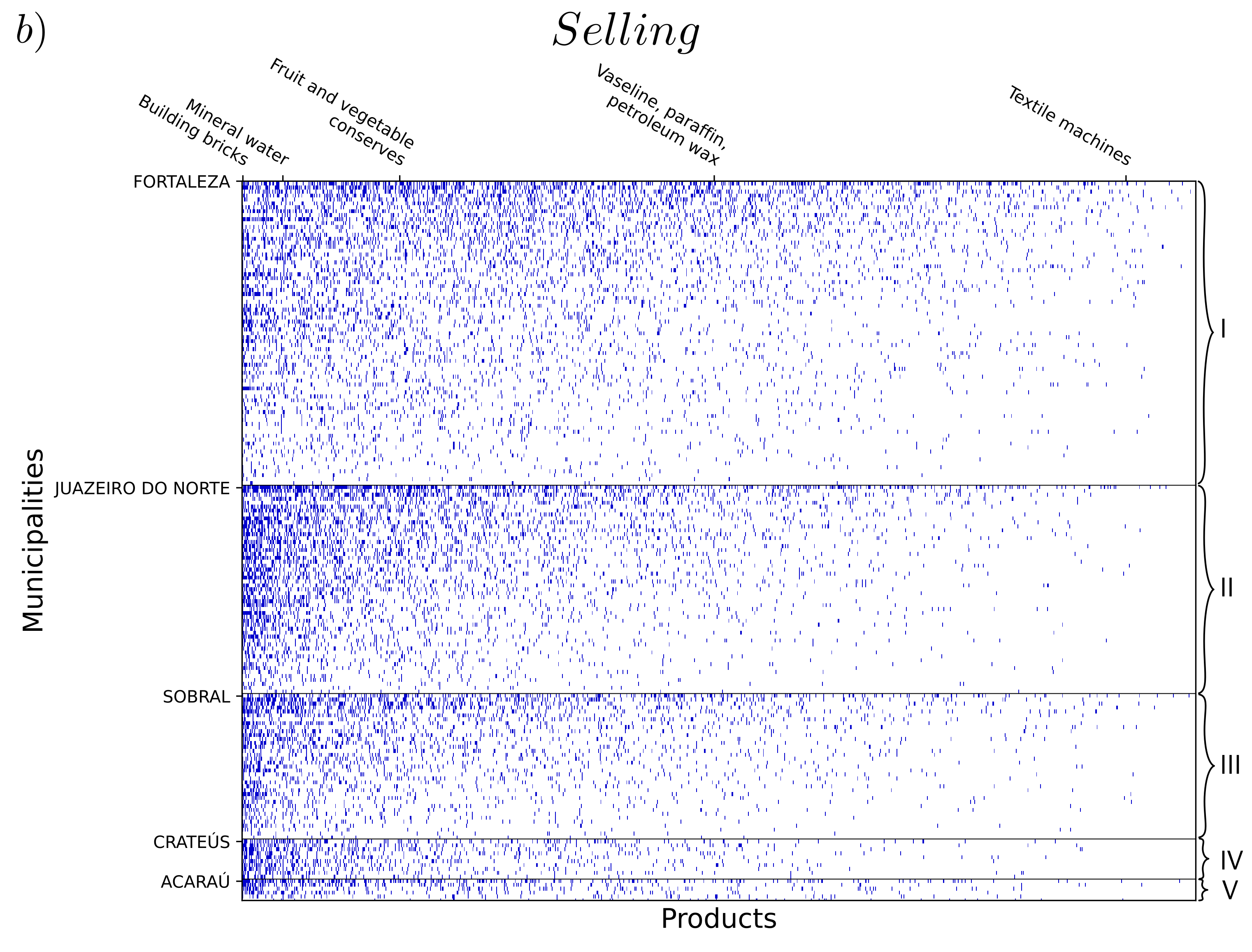}
    \caption{Raster plots of the municipality-product matrices for buying
        \(\boldsymbol{\sigma}^{B}\) (a) and selling \(\boldsymbol{\sigma}^{S}\)
        (b). The rows are sorted in the descending order of the diversity,
        \(D_{c}^{S}=\sum_{p}^{P}\sigma_{c,p}^{S}\), for cities (from top to
        bottom) in each community, and the columns in the descending order of
        the ubiquity, \(U_{p}^{S}=\sum_{c}^{C}\sigma_{c,p}^{S}\), for the
        products (from left to right). Accordingly, for each municipality \(c\),
        the state \(\sigma_{c,p}^{S}\) for the sales  of a product is active
        (\(+1\)) if \(Q_{c,p}^{S}\geq 1\) (blue) or inactive (\(-1\)) if
        \(Q_{c,p}^{S}<1\) (white). The same rule is used for the purchase of a
        product, \(\sigma_{c,p}^{S}\). Highlighted are the leading cities of
        each community in terms of diversity and the most ubiquitous products.
        The products are identified by using a code in the electronic invoices 
        that follows the HS~\cite{ASAKURA1993, wco2017}.}\label{fig:fig04}
\end{figure}

From the RCA matrix for sales, \(\boldsymbol{Q}^{S}\), obtained using
Eq.~(\ref{eq:eq01}), we can construct the corresponding binarized
municipality-product matrices \(\boldsymbol{\sigma}^{S}\), with the
\textit{activity} elements defined by
\begin{equation}
    \sigma_{c,p}^{S} =
    \begin{cases}
        1,  & \mbox{if }Q_{c,p}^{S} \geq 1 \\
        -1, & \mbox{if }Q_{c,p}^{S} < 1,
    \end{cases}
    \label{eq:eq02}
\end{equation}
for each city \(c=1, 2, 3, \ldots , C\) and product \(p=1, 2, 3, \ldots , P\). The same
transformation is applied to the RCA matrix for buying, \(\boldsymbol{Q}^{B}\),
so that the binary matrix \(\boldsymbol{\sigma}^{B}\) with activities elements
\(\sigma_{c,p}^{B}\) can also be obtained. The \textit{diversity} of the basket
of products for sales of a given city \(c\), defined as
\(D_{c}^{S}=\sum_{p}^{P}(\sigma_{c,p}^{S}+1)/2\), corresponds to the number of
relevant products the city sells, while the \textit{ubiquity} for sales of a
given product \(p\), defined as
\(U_{p}^{S}=\sum_{c}^{C}(\sigma_{c,p}^{S}+1)/2\), corresponds to the number of
cities whose baskets of relevant products for sales include the product \(p\).
In the same fashion, the diversity of the basket for buying of a city \(c\),
\(D_{c}^{B}\), and the ubiquity for buying of a product \(p\), \(U_{p}^{B}\), can
be readily obtained from the elements of the binary matrix
\(\boldsymbol{\sigma}^{B}\). 
Figure~\ref{fig:fig04} shows the raster plots of the municipality-product
matrices \(\boldsymbol{\sigma}^{B}\) and \(\boldsymbol{\sigma}^{S}\), with the
cities sorted in the descending order in terms of their respective diversity
(from top to bottom), and the products sorted in the descending order (from left
to right) in terms of their respective ubiquity. 
Upon examination, the municipality-product matrix for buying transactions
(Fig.\ref{fig:fig04}a) is markedly denser than the one for selling transactions
(Fig.\ref{fig:fig04}b). This observation is straightforward to grasp. First notice that
both matrices share the same set of municipalities and the same product
basket. The observed asymmetry reveals the diversification of consumption
behaviors. Specifically, municipalities and their inhabitants engage in purchasing a
wide range of products, many of which are not produced locally. On the other
hand, production trends are far more specialized. Municipalities tend to
focus on manufacturing specific products, influenced by factors such as the
availability of resources, specialized expertise, labor considerations, and
historical contexts. When focussing on commerce, economies of scale become
evident. For some products, distribution is more efficient when undertaken by a
small number of municipalities dealing in significant volumes. Such disparities in
transactional behaviors are responsible for the observed asymmetry.
For the case in study, the most frequent classes of products are ``Appliances
for Agriculture'', ``Rice Cultivation'' and ``Mineral Fuels'' for buying, while
``Building Bricks'', ``Mineral Water'', and ``Fruit and Vegetable Conserves''
are the most frequent classes for selling.
The cities of Fortaleza, Juazeiro do Norte, Maracanaú, Quixadá, and Eusébio are
the cities with the most diverse baskets of products for selling transactions,
while Fortaleza, Juazeiro do Norte, Maracanaú, Sobral, and Caucaia have the
higher diversities in the case of buying.

\section{Maximum Entropy Model}\label{sec:maximum-entropy-model}

In this study we aim to investigate the statistics of product activities of the
cities from the perspective of a pairwise Maximum Entropy Model. 
We follow the approach proposed in Refs.~\cite{Tkacik2009,Bialek2012, Schneidman2006, Tkacik2014, Tkacik2015, Bialek2013,
Bialek2014} to build an Ising-like model for a given community, which is capable
of replicating the average activities and pairwise correlations in terms of the
product baskets for each city in this community. We then evaluate the behavior
of the resulting model in the analogous framework of the corresponding thermal
equilibrium properties.
Accordingly, for a given community, from the observed product series of the
activity of a city \(i\), as defined in Eq.~(\ref{eq:eq02}), we can calculate
its product-average activity as,
\begin{equation}
\avg{\sigma_{i}}^{obs} = \frac{1}{P}\sum_{p=1}^{P}\sigma_{i,p},
\end{equation}
as well as the covariances between the product series of the activities (for
selling or buying) of each pair of cities \(i\) and \(j\),
\begin{equation}
    Cov_{ij}^{obs}= \avg{\sigma_{i} \sigma_{j}}^{obs} - 
    \avg{\sigma_{i}}^{obs} \avg{\sigma_{j}}^{\it{obs}},
\end{equation}
where \({\avg{ \sigma_{i} \sigma_{j} }^{obs}
=\frac{1}{P}\sum_{p=1}^{P}\sigma_{i,p}\sigma_{j,p}}\). Moreover, to model the
observed activities, we consider that \(\sigma\) correspond to Ising-like
variables on a fully connected network of \(C\) sites. Therefore, in analogy
with Statistical Mechanics, \(\ens{ \sigma } = \ens{\sigma_{1,p},\dots,
\sigma_{C,p}}\) would be descriptive of ``the configuration of the community''
with respect to a given product \(p\). The probability distribution \(P(\ens{
\sigma })\) that represents our system is the one that maximizes the entropy,
\(S = -\sum_{\ens{\sigma}} P(\ens{\sigma})\ln{P(\ens{\sigma}   )}\),
while reproducing our observations, \textit{i.e.}, \(\avg{ \sigma_{i} }^{obs}\)
for all \(C\) cities and all \(C(C-1)/2\) values of \(Cov_{ij}^{obs}\).
Given these two additional constraints, it can be readily shown [53] that the
form of \(P(\ens{ \sigma })\) is the Boltzmann’s probability distribution for a
temperature \(T=T_0=1\),
\begin{equation}
    P(\ens{\sigma}) = \frac{1}{Z} e^{-\mathcal{H}(\ens{\sigma})/T},
    \label{eq:boltzmann}
\end{equation}
where \(Z = \sum_{ \ens{\sigma} } e^{-\mathcal{H}( \ens{\sigma})/T} \) is the
partition function and \(\mathcal{H}\) is analogous to a
Hamiltonian with the same form of the Ising model, 
\begin{equation}
    \mathcal{H}(\ens{\sigma}) = -\sum_{i=1}^C h_i \sigma_{i} 
                  - \sum_{i,j > i}^C J_{ij} \sigma_{i} \sigma_{j}.
    \label{eq:hamiltonian}
\end{equation}
It should be noted that this model is derived directly from empirical data
through the maximum entropy principle, rather than being presumed as a
simplified representation of the underlying dynamics. As such, this method
constitutes a precise mapping and not merely a figurative comparison.
Equations~\eqref{eq:boltzmann} and~\eqref{eq:hamiltonian} together are designed to calculate, based on the observed
inter-city correlations, the likelihood of all possible states across the entire
network of cities. This represents a baseline model, implying that the actual
network might exhibit greater complexity than what the maximum entropy model
indicates, but certainly not less.

This Ising-like correspondence naturally leads us to interpret \(h_{i}\) as
the action of a local external stimulus on the product activity of the city
\(i\), analogous to a ``random field'', and \(J_{ij}\) as a ``coupling
coefficient'' between cities \(i\) and \(j\). Such pairwise couplings or
interactions between the product activities of cities give rise to the observed
correlations between them. At this point, we compute the local fields \(h_{i}\)
and the interactions \(J_{ij}\) by directly solving the inverse problem given by
Eq.~(\ref{eq:hamiltonian}). 
The local fields \(h_i\) and interaction constants \(J_{ij}\) are obtained
through the following iterative scheme:
\begin{equation}\label{iterationJ}
    J_{ij} (n+1) = J_{ij} (n) 
    - \eta (n) \left[Cov_{ij}^{MC}-Cov_{ij}^{obs}\right],
\end{equation}
\begin{equation}\label{iterationh}
    h_i (n+1) = h_i (n) 
    - \eta (n) \left[\avg{ \sigma_{i} }^{MC} 
    - \avg{ \sigma_{i} }^{obs}\right],
\end{equation}
\noindent where \(n\) is the iteration parameter and we start with \(n=1\) and
\(h_i(n=1)=0\). The covariance \(Cov_{ij}^{MC}\) between two sites \(i\)
and \(j\) of the Ising-like network of Eq.~(\ref{eq:hamiltonian}) is given by
\(Cov_{ij}^{MC} = \avg{ \sigma_{i} \sigma_{j} }^{MC} -
\avg{ \sigma_{i} }^{MC} \avg{\sigma_{j}}^{MC}\), where the statistical average 
\(\avg{\cdots}^{MC}\) is obtained by performing a Monte Carlo simulation
of the model Eq.~(\ref{eq:hamiltonian}) at temperature \(T_{0} = 1\) using \(h_i
(n)\) and \(J_{ij} (n)\). The function \(\eta (n)\) is a learning rate which
decays like \(1/n^{0.4}\)~\cite{Nguyen2017}. Typically, we iterate till \(n =
80000\). Once we infer the values of \(h_i\) and \(J_{ij}\) that better
reproduce the observed product-average activities \(\left\langle \sigma_{i}
\right\rangle^{obs}\) and covariances \(Cov_{ij}^{obs}\), while
maximizing the entropy, the Boltzmann probability distribution of
Eq.~(\ref{eq:boltzmann}) characterizes the statistics of the product 
activities of the cities composing a given community dataset.

\begin{figure}[ht]
    \includegraphics*[width=\columnwidth]{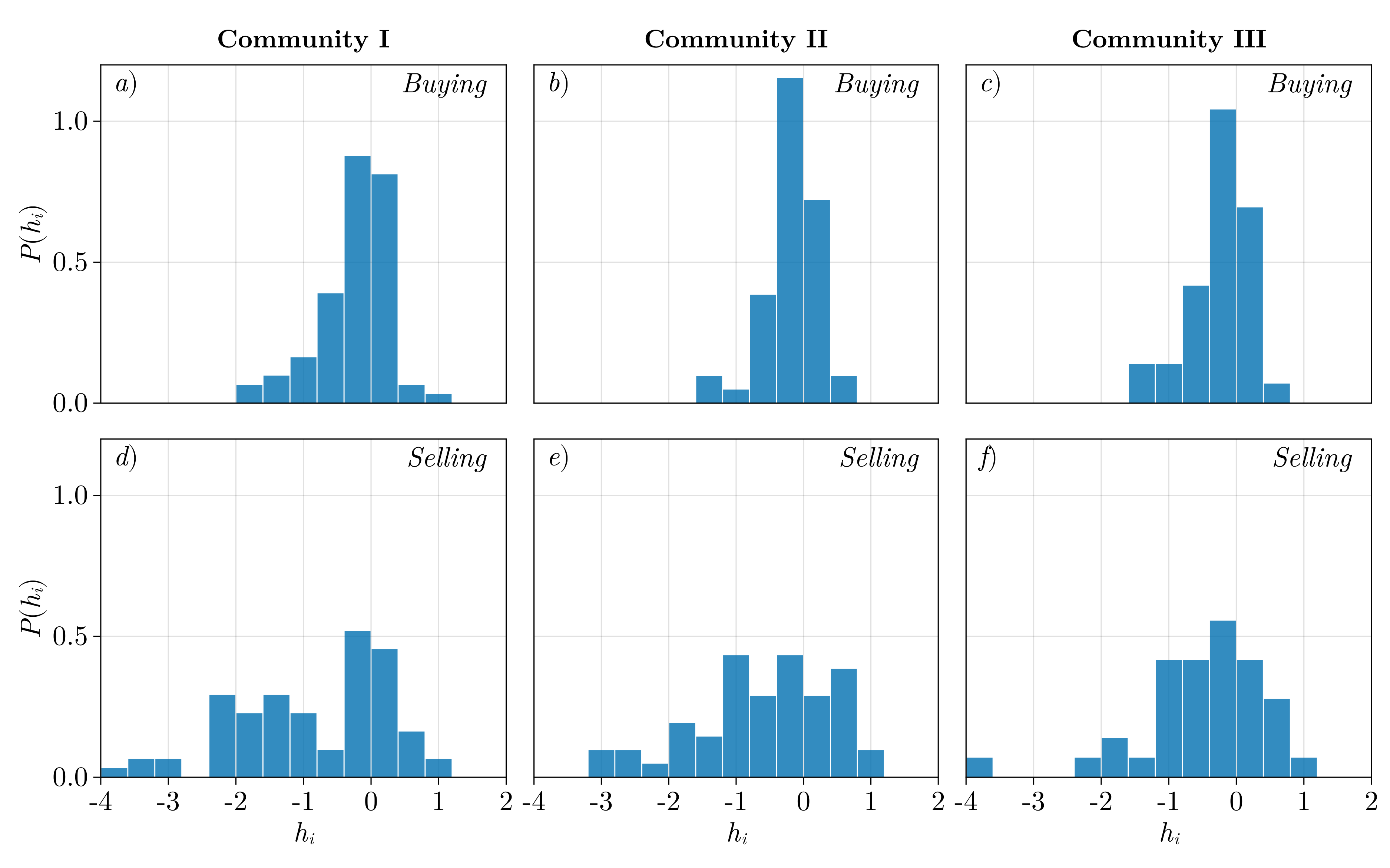}
    \caption{Distribution of local fields \(h_i\) after learning. The figures
        \(a)\), \(b)\), and \(c)\) illustrate the distributions for \textit{selling}
        transactions in Communities I, II, and III, respectively, while figures
        \(d)\), \(e)\), and \(f)\) depict the distributions for \textit{buying}
        transactions in Communities I, II, and III, respectively. It's observed
        that all distributions show negative skewness \(\gamma_1\), indicating a
        predominance of smaller values. The specific values of \(\gamma_1\)
        for each graph are: 
        \(a)\): \(\gamma_1 = -0.90\); 
        \(b)\): \(\gamma_1 = -0.94\);
        \(c)\): \(\gamma_1 = -0.80\); 
        \(d)\): \(\gamma_1 = -0.52\); 
        \(e)\): \(\gamma_1 = -0.63\); 
        \(f)\): \(\gamma_1 = -1.23\).}\label{fig:fig05}
\end{figure}

\begin{figure}[ht]
    \includegraphics*[width=\columnwidth]{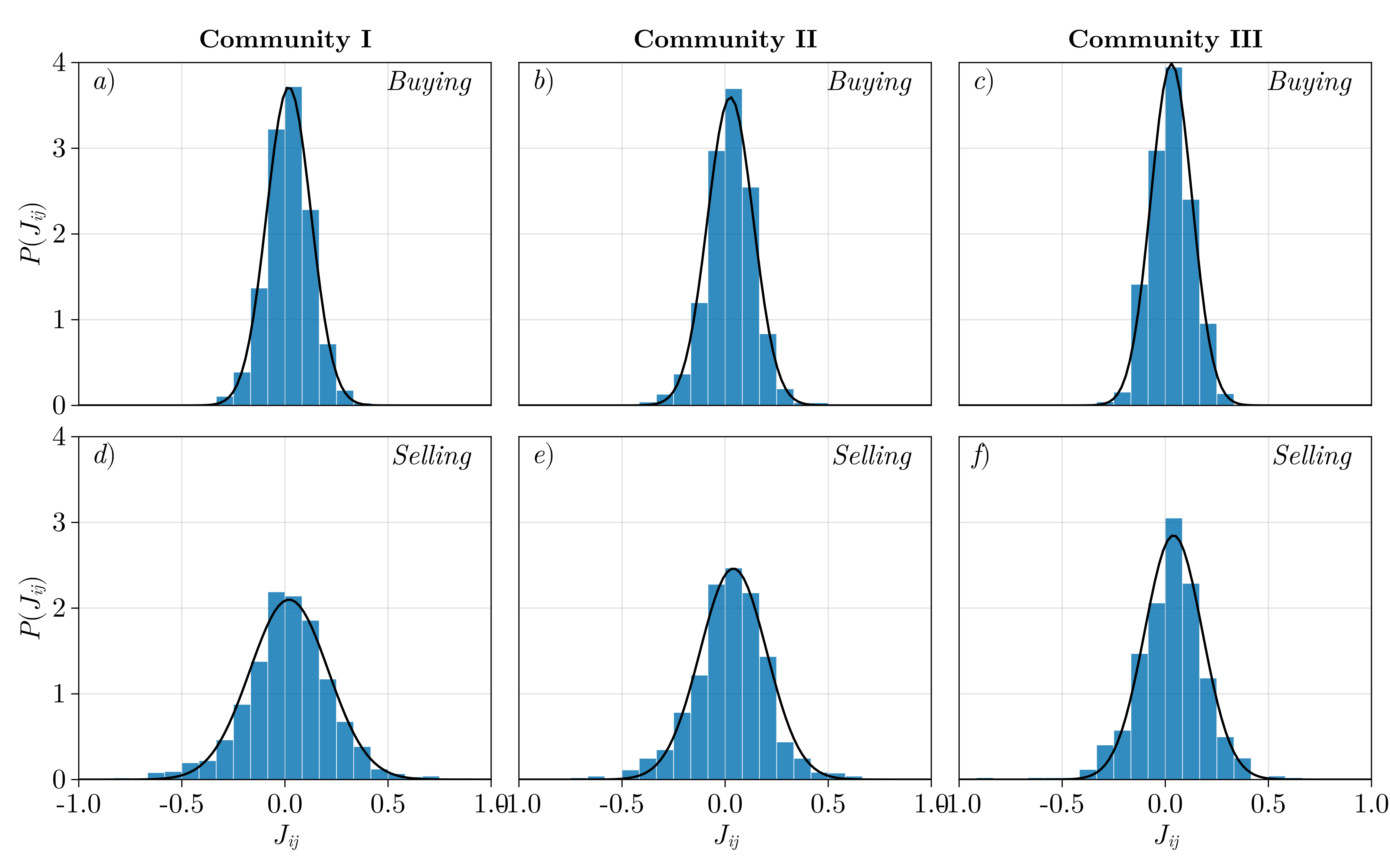}
    \caption{Distributions of couplings \(J_{ij}\). Shown in \(a)\), \(b)\), and
        \(c)\) are the distributions for \textit{selling} transactions 
        and in figures \(d)\), \(e)\), and \(f)\) are the distributions for
        \textit{buying} transactions in Communities I, II, and III,
        respectively. The data were fitted to Gaussian distributions, and the
        adequacy of this model was confirmed by the Kolmogorov-Smirnov (KS)
        test, yielding a \(\text{p-value} > 0.05\) in all cases. This result
        suggests a satisfactory agreement between the observed data and the
        Gaussian model. The fitting parameters (mean \(\mu\) and standard
        deviation \(\sigma\)) and their respective p-values are:  
        \(a)\): \(\mu=0.02\), \(\sigma=0.11\), \(\text{p-value}=0.28\); \(b)\):
        \(\mu=0.03\), \(\sigma=0.12\), \(\text{p-value}=0.21\); \(c)\):
        \(\mu=0.03\), \(\sigma=0.10\), \(\text{p-value}=0.96\); \(d)\):
        \(\mu=0.02\), \(\sigma=0.19\), \(\text{p-value}=0.08\); \(e)\):
        \(\mu=0.04\), \(\sigma=0.16\), \(\text{p-value}=0.10\); \(f)\):
        \(\mu=0.04\), \(\sigma=0.15\), \(\text{p-value}=0.23\);}\label{fig:fig06}
\end{figure}

By solving Eqs.~\eqref{iterationJ} and~\eqref{iterationh} simultaneously, we
obtain for each community and for both buying and selling cases, their
corresponding local fields \(h_i\) and coupling constants \(J_{ij}\). As shown in
Fig.~\ref{fig:fig05}, the distributions of the fields \(h_i\) indicate that they
are predominantly negative, with those for selling transactions being
systematically more negatively skewed than those for buying. Comparatively, as
shown in Fig.~\ref{fig:fig06}, the distributions of the coupling interaction
constants \(J_{ij}\) are symmetrically centered around zero. Additionally, for all
three communities they can be accurately characterized as Gaussian's with mean
values close to zero, but with standard deviations that are systematically
smaller for buying operations than for selling.

\begin{figure}[ht]
    \includegraphics*[width=\columnwidth]{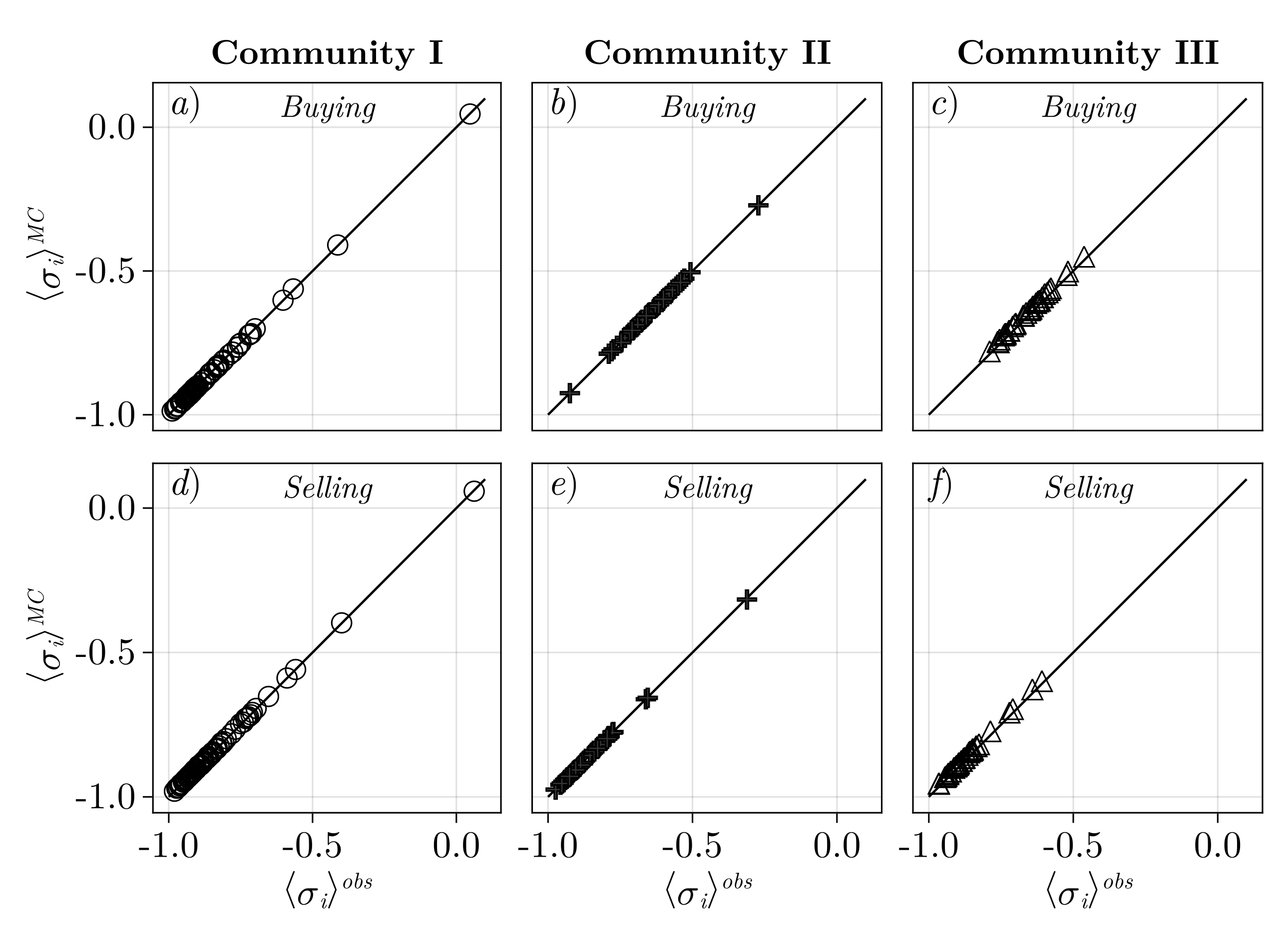}
    \caption{Comparison between the observed average magnetizations, \({\langle
    \sigma_{i}\rangle}^{obs}\), and those predicted by the Maximum Entropy Model
    via Monte Carlo simulations, \({\langle \sigma_{i}\rangle}^{MC}\). 
    Panels \(a)\), \(b)\), and \(c)\) display results corresponding to the
    average selling activities, while panels \(d)\), \(e)\), and \(f)\) pertain
    to buying activities. Each set of three panels represents Communities I, II,
    and III, respectively. The excellent agreement observed across all cases
    highlights the effectiveness of the method in terms of the converged
    parameters \(h_{i}\) and \(J_{ij}\) constituting the Ising-like model, as
    described in Eq.~\eqref{eq:hamiltonian}.}\label{fig:fig07}
\end{figure}

\begin{figure}[ht]
    \includegraphics*[width=\columnwidth]{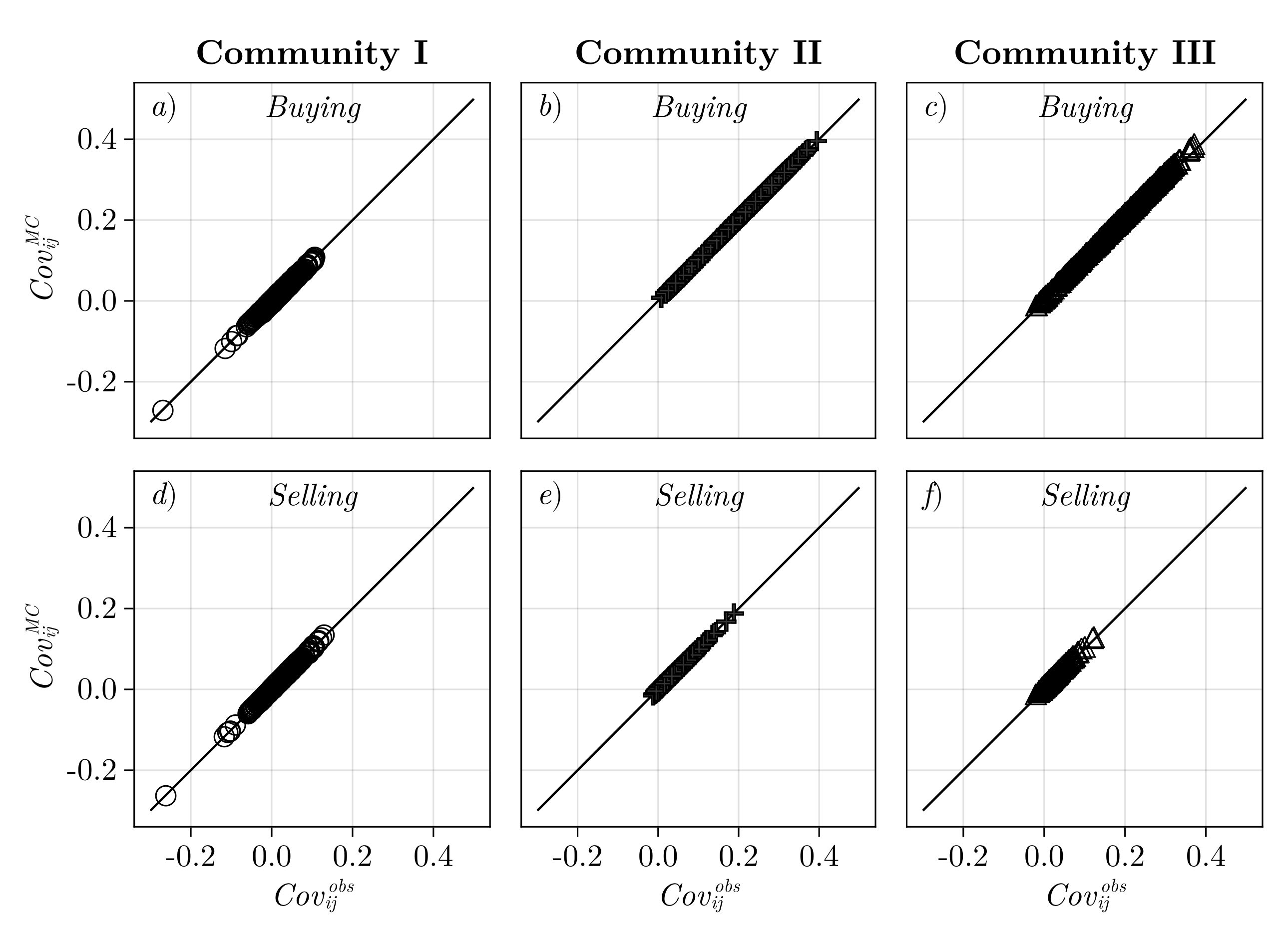}
    \caption{Comparison between observed covariances, \(Cov_{ij}^{obs}\), and
             those predicted by the MEM through Monte Carlo simulations,
             \(Cov_{ij}^{MC}\). The results in \(a)\), \(b)\), and \(c)\)
             correspond to selling average activities, while those in \(d)\),
             \(e)\), and \(f)\) are for buying, with the three panels in both
             sets referring to Communities I, II, and III, respectively. The
             excellent agreement in all cases reflects the convergence of the
             method in terms of the parameters \(h_{i}\) and \(J_{ij}\)
             constituting the Ising-like model
             Eq.~\eqref{eq:hamiltonian}.}\label{fig:fig08}
\end{figure}

\begin{figure}[ht]
    \includegraphics*[width=\columnwidth]{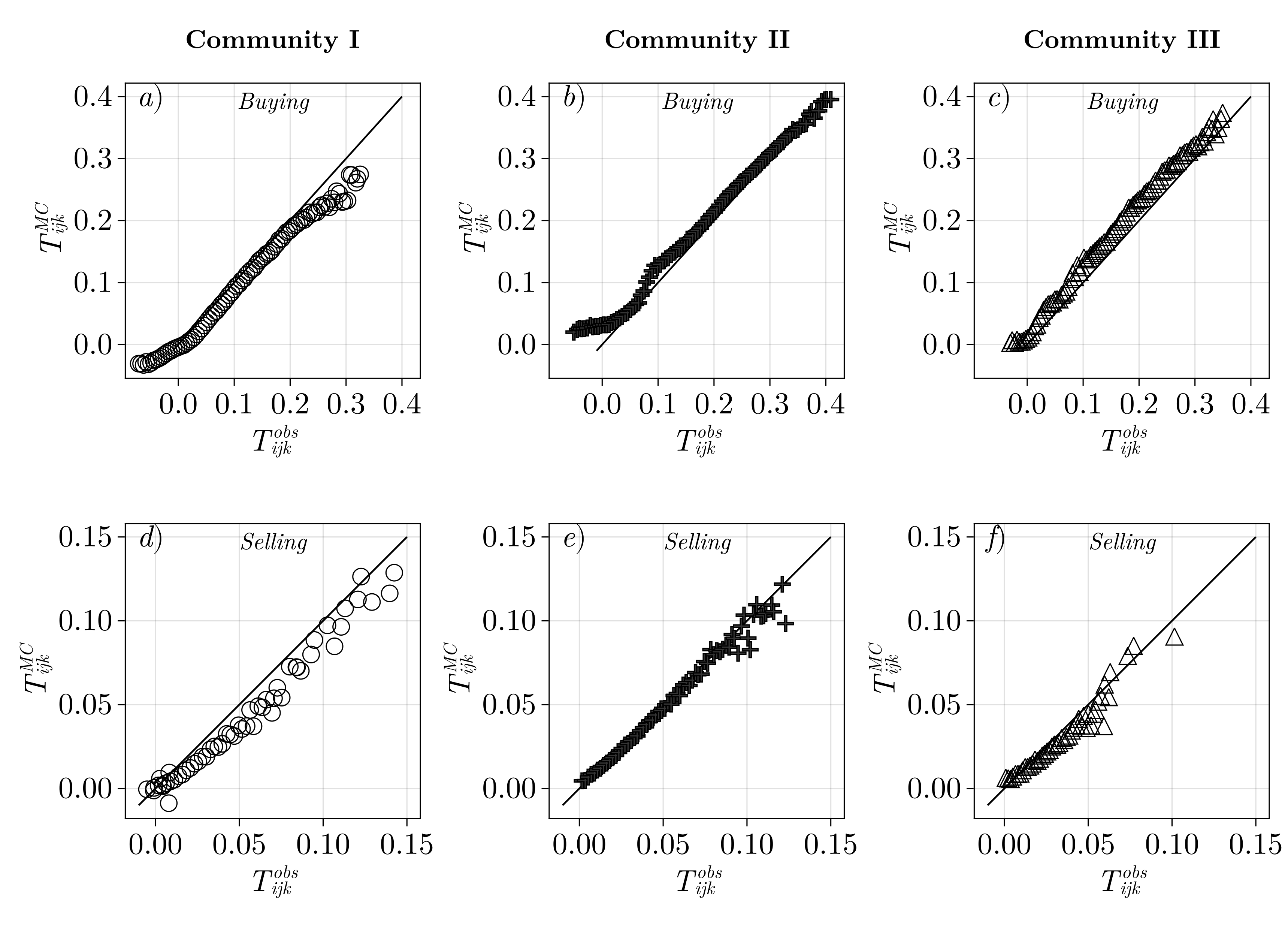}
  \caption{ Comparison of three-point spin correlations as generated by the
            model, \(T^{MC}_{ijk}\), and as observed, \(T^{obs}_{ijk}\).
            Clearly, the model's predictions and the observed triplet
            correlations exhibit a strong correlation in both the buying and
            selling scenarios across all three communities. Furthermore, the
            quantitative alignment between them is also quite satisfactory. The
            relative errors are equal to \(-0.003\), \(-0.001\), and
            \(0.001\) for the buying scenarios and \(-0.005\), \(0.000\),
            and \(-0.001\) for the selling scenarios, in relation to
            Communities I, II, and III, respectively.}\label{fig:fig09}
\end{figure}

To assess the efficacy of the Ising-like model described by
Eq.~(\ref{eq:hamiltonian}) in replicating measured averages derived from
observational data, we compare in Fig.~\ref{fig:fig07} the final
product-averaged magnetizations obtained from the MEM, 
\(\avg{\sigma_{i}}^{MC}\), with their observational counterparts, 
\(\avg{\sigma_{i}}^{obs}\), for each city \(i\) within the three largest
communities labeled I, II, and III, respectively. 
The same is shown in Fig.~\ref{fig:fig08},  but for the covariances
\(Cov_{ij}^{MC}\) against \(Cov_{ij}^{obs}\) between every pair of
cities \(i\) and \(j\) in the aforementioned communities. Clearly, the concordance
between the model-generated and observational metrics is excellent across all
examined cases. 
To provide a more conclusive test of the model, we compared the three-point
activity correlations generated by the model, \(T^{MC}_{ijk}\), with those
observed, \(T^{obs}_{ijk}\), where \(T_{ijk} = \avg{\left(\sigma_i-\avg{\sigma_i}\right) \left(\sigma_j-\avg{\sigma_j}\right) 
\left(\sigma_k-\avg{\sigma_k}\right) }\).
As depicted in Fig.~\ref{fig:fig09}, both the model's predictions and the observed triplet
correlations exhibit a strong correlation across all cases, that is, for both
the buying and selling scenarios in the three communities. 
Moreover, the quantitative alignment between them is also quite satisfactory,
with relative errors, \(\left< \left(T^{MC}_{ijk} -
T^{obs}_{ijk}\right)/T^{obs}_{ijk} \right>\), of 
%
\(-0.33\% \times 10^{-3}\), \(-0.093\%\), and \(0.12\%\) 
for the buying scenarios, and 
%
\(-0.54 \%\), \(0.004\%\), and \(-0.062\%\)
for the selling scenarios, in relation to Communities I, II, and III, respectively.

After demonstrating that the Ising-like model adequately represents the
product-averaged properties of the three largest communities in both selling and buying
scenarios, we explore deeper into the implicit thermodynamic analogy introduced
by the MEM methodology~\cite{Tkacik2014}. Within this framework, using the
learned \(J_{ij}\) and \(h_{i}\) parameters along with Eqs.~\eqref{eq:boltzmann} 
and~\eqref{eq:hamiltonian}, we conduct Monte Carlo simulations at temperatures \(T\)
other than the operational temperature \(T_{0}=1\). 
Given the magnetization for each configuration, defined as
\(M(\{\sigma\})=\left|\frac{1}{C}\sum_{i}^{C}\sigma_{i}\right|\), the order
parameter for a ferromagnetic phase transition can be represented by the
ensemble average at a constant temperature \(T\). This is mathematically
expressed as:

\begin{equation}
    M(T) = \langle M(\{\sigma\}) \rangle_T^{MC}. 
    \label{eq:mag}
\end{equation}
In this equation, the statistical average \(\langle \cdots \rangle_T^{MC}\) is
calculated using Monte Carlo simulations at temperature \(T\). Additionally,
considering the energy per city for each configuration as
\(E(\{\sigma\})=\frac{1}{C}\mathcal{H}(\{\sigma\})\), the
fluctuation-dissipation theorem leads to the following expression for the
specific heat:

\begin{equation} 
    C(T) = \frac{1}{T^{2}} \left( \langle E{(\{\sigma\})}^{2} \rangle - \langle E(\{\sigma\}) \rangle^{2} \right).
    \label{eq:specificHeat}
\end{equation}

In Fig.~\ref{fig:figS1}a and~\ref{fig:figS1}b of the
supplemental material, we present the average
magnetization \(M(T)\) as a function of temperature \(T\) for the buying and
selling activities of the three largest communities, respectively. The
magnetization starts to rapidly decrease near the temperature \(T=1.0\),
indicating a change in behavior that is analogous to a phase transition from
ordered to disordered spin configurations with critical temperatures \(T_c\)
that, for all cases, are close to the operational temperature \(T_{0}\).
The macroscopic ordering diminishes for all analyzed communities as the
temperature ascends. Notably, at temperatures significantly above the critical
ones, the magnetizations observed for the selling curves of all three
communities saturate at larger values compared to the buying cases. This
behavior can be attributed to the local fields distributions showcasing a
pronounced skewness toward negative values in the selling scenario, as shown in
Fig.~\ref{fig:fig05}, thereby mitigating the extent of its order-disorder
transition.

\begin{figure}[t]
    \includegraphics*[width=0.55\columnwidth]{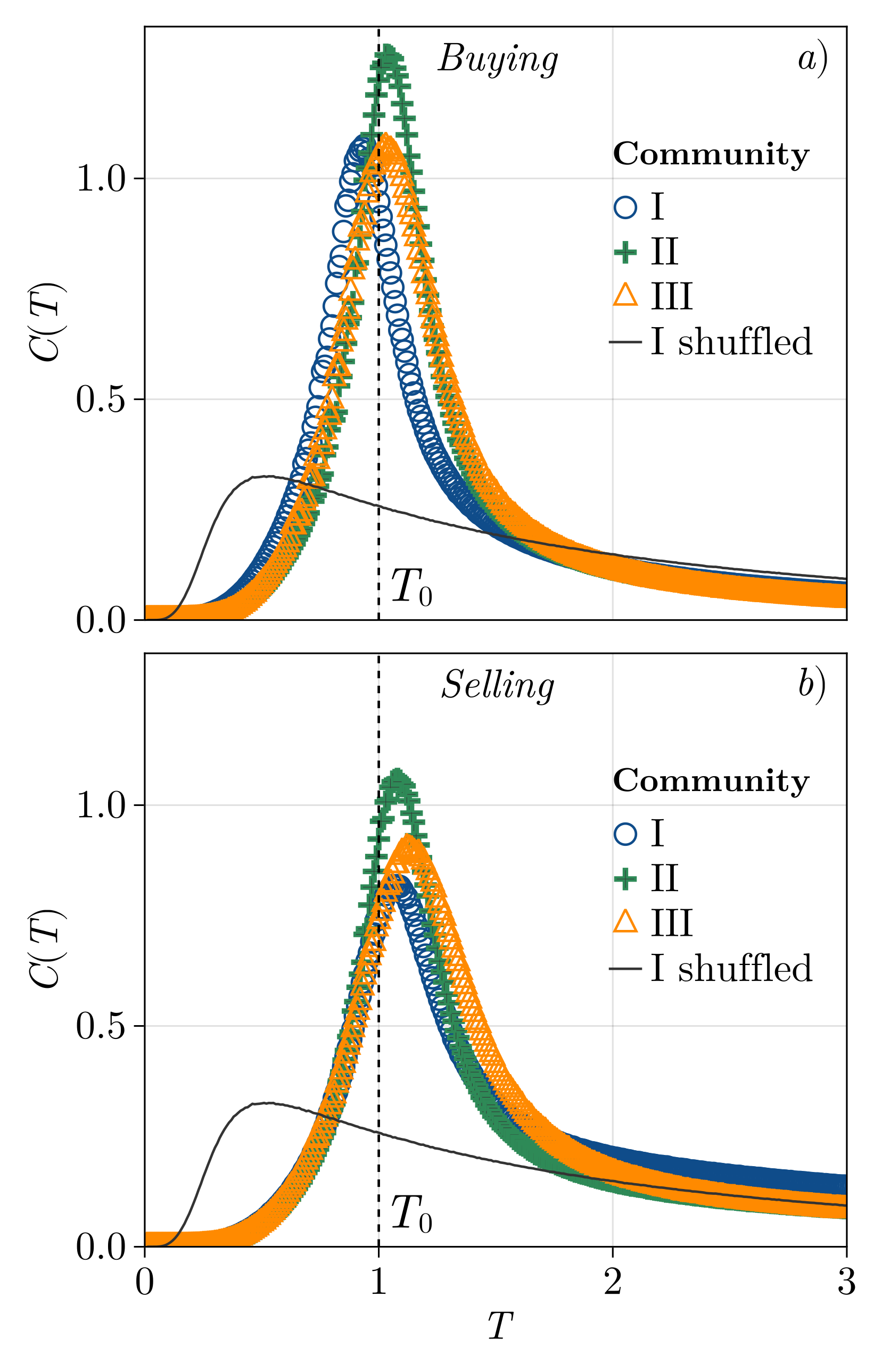}
    \caption{Specific heat \(C(T)\) as a function of the temperature \(T\) for the
             three largest communities of cities in the state. All curves
             exhibit maximal values, \(C^{*}\), at given ``critical temperatures'',
             \(T^{*}\), that are above, but quite  close to the operating
             temperature \(T_{0}=1\). In the case of buying, as shown in \(a)\), the
             peak of the heat capacity becomes closer to the operating point as
             the size of the community decreases. In the case of selling, as
             shown in \(b)\), the larger the community, shorter is its
             corresponding distance to the ``critical point''. In both buying and
             selling cases, however, the maximum value \(C^{*}\) increases with
             system size.}\label{fig:fig10}
\end{figure}

Figures~\ref{fig:fig10}a and~\ref{fig:fig10}b depict the temperature-dependent
specific heat, \( C(T) \), computed for the three largest communities in
relation to their buying and selling product baskets, respectively. In both
scenarios, each curve peaks at a temperature \( T_{c} \), which, while exceeding
the operating temperature \( T_{0}=1 \), remains in close proximity to it (refer
to Table~\ref{tab:tab1} for specific numerical values). Given this observation, it is
reasonable to assert that the learning dynamics of the six Boltzmann machines
predominantly take place within the ``critical region''. This ``critical
temperature'' \( T_{c} \) serves as a threshold, distinguishing the ordered
phases (for \( T < T_{c} \)) from the disordered phases (for \( T > T_{c} \)). A
comparison between Figs.~\ref{fig:fig10}a and \ref{fig:fig10}b indicates that the energy fluctuations for
buying are more pronounced than for selling. Also shown in Figs.~\ref{fig:fig10}a and \ref{fig:fig10}b is
the dependence of the specific heat on temperature, obtained after randomly
shuffling the product series associated with the buying and selling activities
of Community I, respectively. This random shuffling can significantly reduce the
intrinsic correlations present in the sequence of ``spins''. Notably, under these
conditions, the pronounced peak in the specific heat becomes significantly
subdued, effectively undermining the ferromagnetic phase transition originally
observed in the system.

\begin{table}[ht]
    \begin{center}
        \caption{\(\label{tab:tab1} \left(T_{0}-T_{c}\right)\) for different communities 
        and transaction type.} 
        \begin{tabularx}{0.6\textwidth} { 
            >{\raggedright\arraybackslash}X 
            >{\centering\arraybackslash}X 
            >{\centering\arraybackslash}X 
            >{\centering\arraybackslash}X 
            >{\centering\arraybackslash}X 
        }
        \toprule
        & I & II & II & I-shuffled \\   
        \midrule
        Selling & 0.09 & -0.02 & -0.01 & 0.49 \\   
        Buying  & -0.05 & -0.06 & -0.1 & 0.49 \\   
        \bottomrule
        \end{tabularx}
    \end{center}
\end{table}

\FloatBarrier

\section{Conclusions}\label{sec:conclusions}

In summary, we have shown that the directed-weighted trade network among the
municipalities exhibits a nontrivial dependence between the weights of the nodes
on their degrees. From the least-squares fit of the sales inlier data to the
power law, \(W^{S} \sim {\left(k^{S}\right)}^{\beta}\), we obtained \(\beta =
2.07 \pm 0.01\), while the fitting of the purchasing inlier data to \(W^{B} \sim
{\left(k^{B}\right)}^{\beta}\), resulted in \(\beta = 1.77 \pm 0.02\).

In the topological analysis, the use of the Infomap algorithm has revealed five
communities within the trade network among municipalities. Strikingly, the
two-dimensional geographical projections of these modules clearly show that they
are all simply-connected domains. 
This reflects that the trade network is inherently characterized by the flow of
products. This finding highlights the lasting significance of geographic
proximity in shaping trade patterns, even in the era of globalized economies. 
Such contiguous trade communities, identified purely based on trade flow data,
suggest that regional economic interactions are strongly influenced by spatial
factors.
This has significant implications for regional economic policy, emphasizing the
potential for more targeted and regionally coherent economic strategies. 
For comparison, during the writing of this paper, the Ceará state has 14
administrative macro-regions that were established by extensive studies taking
into account technical  criteria related to natural resources, social
solidarity, and polarization around urban centers~\cite{Ataliba2014}.
The contrast between these findings and the results from other algorithms like
the stochastic block model not only reinforces the uniqueness of Infomap's
approach in capturing the flow-based nature of trade but also points to the
nuanced complexities inherent in trade network analysis. 
This understanding can guide future policy decisions and economic planning,
emphasizing the relevance of local and regional contexts in economic networks.

Finally, conceptualizing cities and their trading patterns in terms of buying and selling
products as spin systems has enabled us to harness well-developed theories from
physics to interpret complex economic behaviors. Specifically, for the
purchasing trade operations, we employed a Boltzmann machine that uses a
Hamiltonian analogous to that of a Spin-Glass model, which maximizes the
system's entropy subject to the observed ``magnetizations'' and ``spin-spin
correlations''. Our findings indicate that the system operates very close to a
critical point, poised for a phase transition from ordered states
(``ferromagnetic''), where cities exhibit clustered buying behaviors, to
disordered states (``paramagnetic''), where the decision to buy a given product
appears random. The same analyzes is applicable to selling operations.
Being close to a critical point is particularly revealing, since they are
hallmarked by scale-invariant fluctuations, suggesting that our economic system is on
the verge of a shift, highly sensitive to both external and internal
disruptions.
Furthermore, this proximity to a critical point implies that minor changes in a single city's
economic strategy (or product purchases) could ripple through and potentially
catalyze large-scale alterations in the broader economic landscape. Such a
dynamic signifies a marketplace characterized by rich, interconnected activities
with an inherent potential for rapid evolution. These insights could have
profound implications for understanding the resilience and adaptability of
economies as well as for informing strategic economic policies.

\section{Acknowledgements}

We  thank  the  Brazilian  agencies  CNPq,  CAPES,  FUNCAP and the  National Institute  of  Science  and  Technology  for  Complex  Systems (INCT-SC) for financial support.


\begin{thebibliography}{57}%
\makeatletter
\providecommand \@ifxundefined [1]{%
 \@ifx{#1\undefined}
}%
\providecommand \@ifnum [1]{%
 \ifnum #1\expandafter \@firstoftwo
 \else \expandafter \@secondoftwo
 \fi
}%
\providecommand \@ifx [1]{%
 \ifx #1\expandafter \@firstoftwo
 \else \expandafter \@secondoftwo
 \fi
}%
\providecommand \natexlab [1]{#1}%
\providecommand \enquote  [1]{``#1''}%
\providecommand \bibnamefont  [1]{#1}%
\providecommand \bibfnamefont [1]{#1}%
\providecommand \citenamefont [1]{#1}%
\providecommand \href@noop [0]{\@secondoftwo}%
\providecommand \href [0]{\begingroup \@sanitize@url \@href}%
\providecommand \@href[1]{\@@startlink{#1}\@@href}%
\providecommand \@@href[1]{\endgroup#1\@@endlink}%
\providecommand \@sanitize@url [0]{\catcode `\\12\catcode `\$12\catcode `\&12\catcode `\#12\catcode `\^12\catcode `\_12\catcode `\%12\relax}%
\providecommand \@@startlink[1]{}%
\providecommand \@@endlink[0]{}%
\providecommand \url  [0]{\begingroup\@sanitize@url \@url }%
\providecommand \@url [1]{\endgroup\@href {#1}{\urlprefix }}%
\providecommand \urlprefix  [0]{URL }%
\providecommand \Eprint [0]{\href }%
\providecommand \doibase [0]{https://doi.org/}%
\providecommand \selectlanguage [0]{\@gobble}%
\providecommand \bibinfo  [0]{\@secondoftwo}%
\providecommand \bibfield  [0]{\@secondoftwo}%
\providecommand \translation [1]{[#1]}%
\providecommand \BibitemOpen [0]{}%
\providecommand \bibitemStop [0]{}%
\providecommand \bibitemNoStop [0]{.\EOS\space}%
\providecommand \EOS [0]{\spacefactor3000\relax}%
\providecommand \BibitemShut  [1]{\csname bibitem#1\endcsname}%
\let\auto@bib@innerbib\@empty
\bibitem [{ce_(2023)}]{ce_area}%
  \BibitemOpen
  \href {https://www.ibge.gov.br/cidades-e-estados/ce.html} {\bibinfo {title} {Ceará {\textbar} {Cidades} e {Estados} {\textbar} {IBGE}}} (\bibinfo {year} {2023}),\ \bibinfo {note} {accessed 2023-09-02}\BibitemShut {NoStop}%
\bibitem [{\citenamefont {Hidalgo}\ and\ \citenamefont {Hausmann}(2009)}]{Hidalgo2009}%
  \BibitemOpen
  \bibfield  {author} {\bibinfo {author} {\bibfnamefont {C.~A.}\ \bibnamefont {Hidalgo}}\ and\ \bibinfo {author} {\bibfnamefont {R.}~\bibnamefont {Hausmann}},\ }\bibfield  {title} {\bibinfo {title} {The building blocks of economic complexity},\ }\href {https://doi.org/10.1073/pnas.0900943106} {\bibfield  {journal} {\bibinfo  {journal} {Proc. Natl. Acad. Sci. U. S. A.}\ }\textbf {\bibinfo {volume} {106}},\ \bibinfo {pages} {10570} (\bibinfo {year} {2009})}\BibitemShut {NoStop}%
\bibitem [{\citenamefont {Hidalgo}(2021)}]{Hidalgo2021}%
  \BibitemOpen
  \bibfield  {author} {\bibinfo {author} {\bibfnamefont {C.~A.}\ \bibnamefont {Hidalgo}},\ }\bibfield  {title} {\bibinfo {title} {Economic complexity theory and applications},\ }\href {https://doi.org/10.1038/s42254-020-00275-1} {\bibfield  {journal} {\bibinfo  {journal} {Nat. Rev. Phys.}\ }\textbf {\bibinfo {volume} {3}},\ \bibinfo {pages} {92} (\bibinfo {year} {2021})}\BibitemShut {NoStop}%
\bibitem [{\citenamefont {Tacchella}\ \emph {et~al.}(2012)\citenamefont {Tacchella}, \citenamefont {Cristelli}, \citenamefont {Caldarelli}, \citenamefont {Gabrielli},\ and\ \citenamefont {Pietronero}}]{Tacchella2012}%
  \BibitemOpen
  \bibfield  {author} {\bibinfo {author} {\bibfnamefont {A.}~\bibnamefont {Tacchella}}, \bibinfo {author} {\bibfnamefont {M.}~\bibnamefont {Cristelli}}, \bibinfo {author} {\bibfnamefont {G.}~\bibnamefont {Caldarelli}}, \bibinfo {author} {\bibfnamefont {A.}~\bibnamefont {Gabrielli}},\ and\ \bibinfo {author} {\bibfnamefont {L.}~\bibnamefont {Pietronero}},\ }\bibfield  {title} {\bibinfo {title} {A new metrics for countries' fitness and products' complexity},\ }\href {https://doi.org/10.1038/srep00723} {\bibfield  {journal} {\bibinfo  {journal} {Sci. Rep.}\ }\textbf {\bibinfo {volume} {2}},\ \bibinfo {pages} {723} (\bibinfo {year} {2012})}\BibitemShut {NoStop}%
\bibitem [{\citenamefont {Tacchella}\ \emph {et~al.}(2013)\citenamefont {Tacchella}, \citenamefont {Cristelli}, \citenamefont {Caldarelli}, \citenamefont {Gabrielli},\ and\ \citenamefont {Pietronero}}]{Tacchella2013}%
  \BibitemOpen
  \bibfield  {author} {\bibinfo {author} {\bibfnamefont {A.}~\bibnamefont {Tacchella}}, \bibinfo {author} {\bibfnamefont {M.}~\bibnamefont {Cristelli}}, \bibinfo {author} {\bibfnamefont {G.}~\bibnamefont {Caldarelli}}, \bibinfo {author} {\bibfnamefont {A.}~\bibnamefont {Gabrielli}},\ and\ \bibinfo {author} {\bibfnamefont {L.}~\bibnamefont {Pietronero}},\ }\bibfield  {title} {\bibinfo {title} {Economic complexity: Conceptual grounding of a new metrics for global competitiveness},\ }\href {https://doi.org/10.1016/j.jedc.2013.04.006} {\bibfield  {journal} {\bibinfo  {journal} {J. Econ. Dynam. Control}\ }\textbf {\bibinfo {volume} {37}},\ \bibinfo {pages} {1683} (\bibinfo {year} {2013})}\BibitemShut {NoStop}%
\bibitem [{\citenamefont {Hidalgo}\ \emph {et~al.}(2007)\citenamefont {Hidalgo}, \citenamefont {Klinger}, \citenamefont {Barabasi},\ and\ \citenamefont {Hausmann}}]{Hidalgo2007}%
  \BibitemOpen
  \bibfield  {author} {\bibinfo {author} {\bibfnamefont {C.~A.}\ \bibnamefont {Hidalgo}}, \bibinfo {author} {\bibfnamefont {B.}~\bibnamefont {Klinger}}, \bibinfo {author} {\bibfnamefont {A.-L.}\ \bibnamefont {Barabasi}},\ and\ \bibinfo {author} {\bibfnamefont {R.}~\bibnamefont {Hausmann}},\ }\bibfield  {title} {\bibinfo {title} {The product space conditions the development of nations},\ }\href {https://doi.org/10.1126/science.1144581} {\bibfield  {journal} {\bibinfo  {journal} {Science}\ }\textbf {\bibinfo {volume} {317}},\ \bibinfo {pages} {482} (\bibinfo {year} {2007})}\BibitemShut {NoStop}%
\bibitem [{\citenamefont {Balassa}(1965)}]{Balassa1965}%
  \BibitemOpen
  \bibfield  {author} {\bibinfo {author} {\bibfnamefont {B.}~\bibnamefont {Balassa}},\ }\bibfield  {title} {\bibinfo {title} {Trade liberalisation and “revealed” comparative advantage},\ }\href {https://doi.org/10.1111/j.1467-9957.1965.tb00050.x} {\bibfield  {journal} {\bibinfo  {journal} {Manchester Sch.}\ }\textbf {\bibinfo {volume} {33}},\ \bibinfo {pages} {99} (\bibinfo {year} {1965})}\BibitemShut {NoStop}%
\bibitem [{\citenamefont {Balland}\ \emph {et~al.}(2022)\citenamefont {Balland}, \citenamefont {Broekel}, \citenamefont {Diodato}, \citenamefont {Giuliani}, \citenamefont {Hausmann}, \citenamefont {O'Clery},\ and\ \citenamefont {Rigby}}]{Balland2022}%
  \BibitemOpen
  \bibfield  {author} {\bibinfo {author} {\bibfnamefont {P.-A.}\ \bibnamefont {Balland}}, \bibinfo {author} {\bibfnamefont {T.}~\bibnamefont {Broekel}}, \bibinfo {author} {\bibfnamefont {D.}~\bibnamefont {Diodato}}, \bibinfo {author} {\bibfnamefont {E.}~\bibnamefont {Giuliani}}, \bibinfo {author} {\bibfnamefont {R.}~\bibnamefont {Hausmann}}, \bibinfo {author} {\bibfnamefont {N.}~\bibnamefont {O'Clery}},\ and\ \bibinfo {author} {\bibfnamefont {D.}~\bibnamefont {Rigby}},\ }\bibfield  {title} {\bibinfo {title} {The new paradigm of economic complexity},\ }\href {https://doi.org/10.1016/j.respol.2021.104450} {\bibfield  {journal} {\bibinfo  {journal} {Res. Pol.}\ }\textbf {\bibinfo {volume} {51}},\ \bibinfo {pages} {104450} (\bibinfo {year} {2022})}\BibitemShut {NoStop}%
\bibitem [{\citenamefont {Brummitt}\ \emph {et~al.}(2020)\citenamefont {Brummitt}, \citenamefont {Gómez-Liévano}, \citenamefont {Hausmann},\ and\ \citenamefont {Bonds}}]{Brummitt2020}%
  \BibitemOpen
  \bibfield  {author} {\bibinfo {author} {\bibfnamefont {C.~D.}\ \bibnamefont {Brummitt}}, \bibinfo {author} {\bibfnamefont {A.}~\bibnamefont {Gómez-Liévano}}, \bibinfo {author} {\bibfnamefont {R.}~\bibnamefont {Hausmann}},\ and\ \bibinfo {author} {\bibfnamefont {M.~H.}\ \bibnamefont {Bonds}},\ }\bibfield  {title} {\bibinfo {title} {Machine-learned patterns suggest that diversification drives economic development},\ }\href {https://doi.org/10.1098/rsif.2019.0283} {\bibfield  {journal} {\bibinfo  {journal} {J. R. Soc. Interface}\ }\textbf {\bibinfo {volume} {17}},\ \bibinfo {pages} {20190283} (\bibinfo {year} {2020})}\BibitemShut {NoStop}%
\bibitem [{\citenamefont {Albora}\ \emph {et~al.}(2023)\citenamefont {Albora}, \citenamefont {Pietronero}, \citenamefont {Tacchella},\ and\ \citenamefont {Zaccaria}}]{Albora2023}%
  \BibitemOpen
  \bibfield  {author} {\bibinfo {author} {\bibfnamefont {G.}~\bibnamefont {Albora}}, \bibinfo {author} {\bibfnamefont {L.}~\bibnamefont {Pietronero}}, \bibinfo {author} {\bibfnamefont {A.}~\bibnamefont {Tacchella}},\ and\ \bibinfo {author} {\bibfnamefont {A.}~\bibnamefont {Zaccaria}},\ }\bibfield  {title} {\bibinfo {title} {Product progression: a machine learning approach to forecasting industrial upgrading},\ }\href {https://doi.org/10.1038/s41598-023-28179-x} {\bibfield  {journal} {\bibinfo  {journal} {Sci. Rep.}\ }\textbf {\bibinfo {volume} {13}},\ \bibinfo {pages} {1421} (\bibinfo {year} {2023})}\BibitemShut {NoStop}%
\bibitem [{\citenamefont {Asakura}(1993)}]{ASAKURA1993}%
  \BibitemOpen
  \bibfield  {author} {\bibinfo {author} {\bibfnamefont {H.}~\bibnamefont {Asakura}},\ }\bibfield  {title} {\bibinfo {title} {The harmonized system and rules of origin},\ }\href@noop {} {\bibfield  {journal} {\bibinfo  {journal} {J. World Trade}\ }\textbf {\bibinfo {volume} {27}},\ \bibinfo {pages} {5} (\bibinfo {year} {1993})}\BibitemShut {NoStop}%
\bibitem [{\citenamefont {Organization}(2017)}]{wco2017}%
  \BibitemOpen
  \bibfield  {author} {\bibinfo {author} {\bibfnamefont {W.~C.}\ \bibnamefont {Organization}},\ }\href {https://www.wcoomd.org/en/topics/nomenclature/instrument-and-tools/hs-nomenclature-2017-edition.aspx} {\bibinfo {title} {Hs nomenclature 2017 edition}} (\bibinfo {year} {2017}),\ \bibinfo {note} {accessed 2023-05-12}\BibitemShut {NoStop}%
\bibitem [{\citenamefont {Jean}\ \emph {et~al.}(2016)\citenamefont {Jean}, \citenamefont {Burke}, \citenamefont {Xie}, \citenamefont {Davis}, \citenamefont {Lobell},\ and\ \citenamefont {Ermon}}]{Jean2016}%
  \BibitemOpen
  \bibfield  {author} {\bibinfo {author} {\bibfnamefont {N.}~\bibnamefont {Jean}}, \bibinfo {author} {\bibfnamefont {M.}~\bibnamefont {Burke}}, \bibinfo {author} {\bibfnamefont {M.}~\bibnamefont {Xie}}, \bibinfo {author} {\bibfnamefont {W.~M.}\ \bibnamefont {Davis}}, \bibinfo {author} {\bibfnamefont {D.~B.}\ \bibnamefont {Lobell}},\ and\ \bibinfo {author} {\bibfnamefont {S.}~\bibnamefont {Ermon}},\ }\bibfield  {title} {\bibinfo {title} {Combining satellite imagery and machine learning to predict poverty},\ }\href {https://doi.org/10.1126/science.aaf7894} {\bibfield  {journal} {\bibinfo  {journal} {Science}\ }\textbf {\bibinfo {volume} {353}},\ \bibinfo {pages} {790} (\bibinfo {year} {2016})}\BibitemShut {NoStop}%
\bibitem [{\citenamefont {Operti}\ \emph {et~al.}(2018)\citenamefont {Operti}, \citenamefont {Pugliese}, \citenamefont {Andrade}, \citenamefont {Pietronero},\ and\ \citenamefont {Gabrielli}}]{Operti2018}%
  \BibitemOpen
  \bibfield  {author} {\bibinfo {author} {\bibfnamefont {F.~G.}\ \bibnamefont {Operti}}, \bibinfo {author} {\bibfnamefont {E.}~\bibnamefont {Pugliese}}, \bibinfo {author} {\bibfnamefont {J.~S.}\ \bibnamefont {Andrade}}, \bibinfo {author} {\bibfnamefont {L.}~\bibnamefont {Pietronero}},\ and\ \bibinfo {author} {\bibfnamefont {A.}~\bibnamefont {Gabrielli}},\ }\bibfield  {title} {\bibinfo {title} {Dynamics in the fitness-income plane: Brazilian states vs world countries},\ }\href {https://doi.org/10.1371/journal.pone.0197616} {\bibfield  {journal} {\bibinfo  {journal} {PLoS ONE}\ }\textbf {\bibinfo {volume} {13}},\ \bibinfo {pages} {e0217034} (\bibinfo {year} {2018})}\BibitemShut {NoStop}%
\bibitem [{\citenamefont {Cristelli}\ \emph {et~al.}(2013)\citenamefont {Cristelli}, \citenamefont {Gabrielli}, \citenamefont {Tacchella}, \citenamefont {Caldarelli},\ and\ \citenamefont {Pietronero}}]{Cristelli2013}%
  \BibitemOpen
  \bibfield  {author} {\bibinfo {author} {\bibfnamefont {M.}~\bibnamefont {Cristelli}}, \bibinfo {author} {\bibfnamefont {A.}~\bibnamefont {Gabrielli}}, \bibinfo {author} {\bibfnamefont {A.}~\bibnamefont {Tacchella}}, \bibinfo {author} {\bibfnamefont {G.}~\bibnamefont {Caldarelli}},\ and\ \bibinfo {author} {\bibfnamefont {L.}~\bibnamefont {Pietronero}},\ }\bibfield  {title} {\bibinfo {title} {Measuring the intangibles: A metrics for the economic complexity of countries and products},\ }\bibfield  {journal} {\bibinfo  {journal} {PLoS ONE}\ }\textbf {\bibinfo {volume} {8}},\ \href {https://doi.org/10.1371/journal.pone.0070726} {10.1371/journal.pone.0070726} (\bibinfo {year} {2013})\BibitemShut {NoStop}%
\bibitem [{\citenamefont {Girvan}\ and\ \citenamefont {Newman}(2002)}]{Girvan2002}%
  \BibitemOpen
  \bibfield  {author} {\bibinfo {author} {\bibfnamefont {M.}~\bibnamefont {Girvan}}\ and\ \bibinfo {author} {\bibfnamefont {M.~E.~J.}\ \bibnamefont {Newman}},\ }\bibfield  {title} {\bibinfo {title} {Community structure in social and biological networks},\ }\href {https://doi.org/10.1073/pnas.122653799} {\bibfield  {journal} {\bibinfo  {journal} {Proc. Natl. Acad. Sci. U. S. A.}\ }\textbf {\bibinfo {volume} {99}},\ \bibinfo {pages} {7821} (\bibinfo {year} {2002})}\BibitemShut {NoStop}%
\bibitem [{\citenamefont {Bathelt}\ \emph {et~al.}(2004)\citenamefont {Bathelt}, \citenamefont {Malmberg},\ and\ \citenamefont {Maskell}}]{Bathelt2004}%
  \BibitemOpen
  \bibfield  {author} {\bibinfo {author} {\bibfnamefont {H.}~\bibnamefont {Bathelt}}, \bibinfo {author} {\bibfnamefont {A.}~\bibnamefont {Malmberg}},\ and\ \bibinfo {author} {\bibfnamefont {P.}~\bibnamefont {Maskell}},\ }\bibfield  {title} {\bibinfo {title} {Clusters and knowledge: local buzz, global pipelines and the process of knowledge creation},\ }\href {https://doi.org/10.1191/0309132504ph469oa} {\bibfield  {journal} {\bibinfo  {journal} {Prog. Hum. Geog.}\ }\textbf {\bibinfo {volume} {28}},\ \bibinfo {pages} {31} (\bibinfo {year} {2004})}\BibitemShut {NoStop}%
\bibitem [{\citenamefont {Peixoto}(2014)}]{Peixoto2014}%
  \BibitemOpen
  \bibfield  {author} {\bibinfo {author} {\bibfnamefont {T.~P.}\ \bibnamefont {Peixoto}},\ }\bibfield  {title} {\bibinfo {title} {Efficient monte carlo and greedy heuristic for the inference of stochastic block models},\ }\href {https://doi.org/10.1103/PhysRevE.89.012804} {\bibfield  {journal} {\bibinfo  {journal} {Phys. Rev. E}\ }\textbf {\bibinfo {volume} {89}},\ \bibinfo {pages} {012804} (\bibinfo {year} {2014})}\BibitemShut {NoStop}%
\bibitem [{\citenamefont {Bialek}(2012)}]{bialek2012biophysics}%
  \BibitemOpen
  \bibfield  {author} {\bibinfo {author} {\bibfnamefont {W.}~\bibnamefont {Bialek}},\ }\href@noop {} {\emph {\bibinfo {title} {Biophysics: searching for principles}}}\ (\bibinfo  {publisher} {Princeton University Press},\ \bibinfo {year} {2012})\BibitemShut {NoStop}%
\bibitem [{\citenamefont {Cocco}\ \emph {et~al.}(2009)\citenamefont {Cocco}, \citenamefont {Leibler},\ and\ \citenamefont {Monasson}}]{Cocco2009}%
  \BibitemOpen
  \bibfield  {author} {\bibinfo {author} {\bibfnamefont {S.}~\bibnamefont {Cocco}}, \bibinfo {author} {\bibfnamefont {S.}~\bibnamefont {Leibler}},\ and\ \bibinfo {author} {\bibfnamefont {R.}~\bibnamefont {Monasson}},\ }\bibfield  {title} {\bibinfo {title} {Neuronal couplings between retinal ganglion cells inferred by efficient inverse statistical physics methods},\ }\href {https://doi.org/10.1073/PNAS.0906705106/SUPPL_FILE/APPENDIX_PDF.PDF} {\bibfield  {journal} {\bibinfo  {journal} {Proc. Natl. Acad. Sci. U. S. A.}\ }\textbf {\bibinfo {volume} {106}},\ \bibinfo {pages} {14058} (\bibinfo {year} {2009})}\BibitemShut {NoStop}%
\bibitem [{\citenamefont {Tkacik}\ \emph {et~al.}(2009)\citenamefont {Tkacik}, \citenamefont {Schneidman}, \citenamefont {au2},\ and\ \citenamefont {Bialek}}]{Tkacik2009}%
  \BibitemOpen
  \bibfield  {author} {\bibinfo {author} {\bibfnamefont {G.}~\bibnamefont {Tkacik}}, \bibinfo {author} {\bibfnamefont {E.}~\bibnamefont {Schneidman}}, \bibinfo {author} {\bibfnamefont {M.~J. B.~I.}\ \bibnamefont {au2}},\ and\ \bibinfo {author} {\bibfnamefont {W.}~\bibnamefont {Bialek}},\ }\href {http://arxiv.org/abs/0912.5409} {\bibinfo {title} {Spin glass models for a network of real neurons}} (\bibinfo {year} {2009}),\ \Eprint {https://arxiv.org/abs/0912.5409} {arXiv:0912.5409 [q-bio.NC]} \BibitemShut {NoStop}%
\bibitem [{\citenamefont {Shlens}\ \emph {et~al.}(2006)\citenamefont {Shlens}, \citenamefont {Field}, \citenamefont {Gauthier}, \citenamefont {Grivich}, \citenamefont {Petrusca}, \citenamefont {Sher}, \citenamefont {Litke},\ and\ \citenamefont {Chichilnisky}}]{Shlens2006}%
  \BibitemOpen
  \bibfield  {author} {\bibinfo {author} {\bibfnamefont {J.}~\bibnamefont {Shlens}}, \bibinfo {author} {\bibfnamefont {G.~D.}\ \bibnamefont {Field}}, \bibinfo {author} {\bibfnamefont {J.~L.}\ \bibnamefont {Gauthier}}, \bibinfo {author} {\bibfnamefont {M.~I.}\ \bibnamefont {Grivich}}, \bibinfo {author} {\bibfnamefont {D.}~\bibnamefont {Petrusca}}, \bibinfo {author} {\bibfnamefont {A.}~\bibnamefont {Sher}}, \bibinfo {author} {\bibfnamefont {A.~M.}\ \bibnamefont {Litke}},\ and\ \bibinfo {author} {\bibfnamefont {E.~J.}\ \bibnamefont {Chichilnisky}},\ }\bibfield  {title} {\bibinfo {title} {The structure of multi-neuron firing patterns in primate retina},\ }\href {https://doi.org/10.1523/JNEUROSCI.1282-06.2006} {\bibfield  {journal} {\bibinfo  {journal} {J. Neurosci.}\ }\textbf {\bibinfo {volume} {26}},\ \bibinfo {pages} {8254} (\bibinfo {year} {2006})}\BibitemShut {NoStop}%
\bibitem [{\citenamefont {Tang}\ \emph {et~al.}(2008)\citenamefont {Tang}, \citenamefont {Jackson}, \citenamefont {Hobbs}, \citenamefont {Chen}, \citenamefont {Smith}, \citenamefont {Patel}, \citenamefont {Prieto}, \citenamefont {Petrusca}, \citenamefont {Grivich}, \citenamefont {Sher}, \citenamefont {Hottowy}, \citenamefont {Dabrowski}, \citenamefont {Litke},\ and\ \citenamefont {Beggs}}]{tang2008maximum}%
  \BibitemOpen
  \bibfield  {author} {\bibinfo {author} {\bibfnamefont {A.}~\bibnamefont {Tang}}, \bibinfo {author} {\bibfnamefont {D.}~\bibnamefont {Jackson}}, \bibinfo {author} {\bibfnamefont {J.}~\bibnamefont {Hobbs}}, \bibinfo {author} {\bibfnamefont {W.}~\bibnamefont {Chen}}, \bibinfo {author} {\bibfnamefont {J.~L.}\ \bibnamefont {Smith}}, \bibinfo {author} {\bibfnamefont {H.}~\bibnamefont {Patel}}, \bibinfo {author} {\bibfnamefont {A.}~\bibnamefont {Prieto}}, \bibinfo {author} {\bibfnamefont {D.}~\bibnamefont {Petrusca}}, \bibinfo {author} {\bibfnamefont {M.~I.}\ \bibnamefont {Grivich}}, \bibinfo {author} {\bibfnamefont {A.}~\bibnamefont {Sher}}, \bibinfo {author} {\bibfnamefont {P.}~\bibnamefont {Hottowy}}, \bibinfo {author} {\bibfnamefont {W.}~\bibnamefont {Dabrowski}}, \bibinfo {author} {\bibfnamefont {A.~M.}\ \bibnamefont {Litke}},\ and\ \bibinfo {author} {\bibfnamefont {J.~M.}\ \bibnamefont {Beggs}},\ }\bibfield  {title} {\bibinfo {title} {A maximum entropy model applied to spatial and temporal correlations from cortical networks in vitro},\ }\href {https://doi.org/10.1523/JNEUROSCI.3359-07.2008} {\bibfield  {journal} {\bibinfo  {journal} {J. Neurosci.}\ }\textbf {\bibinfo {volume} {28}},\ \bibinfo {pages} {505} (\bibinfo {year} {2008})}\BibitemShut {NoStop}%
\bibitem [{\citenamefont {Mora}\ \emph {et~al.}(2015)\citenamefont {Mora}, \citenamefont {Deny},\ and\ \citenamefont {Marre}}]{Mora2005}%
  \BibitemOpen
  \bibfield  {author} {\bibinfo {author} {\bibfnamefont {T.}~\bibnamefont {Mora}}, \bibinfo {author} {\bibfnamefont {S.}~\bibnamefont {Deny}},\ and\ \bibinfo {author} {\bibfnamefont {O.}~\bibnamefont {Marre}},\ }\bibfield  {title} {\bibinfo {title} {Dynamical criticality in the collective activity of a population of retinal neurons},\ }\href {https://doi.org/10.1103/PHYSREVLETT.114.078105/FIGURES/3/MEDIUM} {\bibfield  {journal} {\bibinfo  {journal} {Phys. Rev. Lett.}\ }\textbf {\bibinfo {volume} {114}},\ \bibinfo {pages} {078105} (\bibinfo {year} {2015})}\BibitemShut {NoStop}%
\bibitem [{\citenamefont {Lotfi}\ \emph {et~al.}(2020)\citenamefont {Lotfi}, \citenamefont {Fontenele}, \citenamefont {Feliciano}, \citenamefont {Aguiar}, \citenamefont {Vasconcelos}, \citenamefont {Soares-Cunha}, \citenamefont {Coimbra}, \citenamefont {Rodrigues}, \citenamefont {Sousa}, \citenamefont {Copelli},\ and\ \citenamefont {Carelli}}]{Lotfi2020}%
  \BibitemOpen
  \bibfield  {author} {\bibinfo {author} {\bibfnamefont {N.}~\bibnamefont {Lotfi}}, \bibinfo {author} {\bibfnamefont {A.~J.}\ \bibnamefont {Fontenele}}, \bibinfo {author} {\bibfnamefont {T.}~\bibnamefont {Feliciano}}, \bibinfo {author} {\bibfnamefont {L.~A.}\ \bibnamefont {Aguiar}}, \bibinfo {author} {\bibfnamefont {N.~A.~D.}\ \bibnamefont {Vasconcelos}}, \bibinfo {author} {\bibfnamefont {C.}~\bibnamefont {Soares-Cunha}}, \bibinfo {author} {\bibfnamefont {B.}~\bibnamefont {Coimbra}}, \bibinfo {author} {\bibfnamefont {A.~J.}\ \bibnamefont {Rodrigues}}, \bibinfo {author} {\bibfnamefont {N.}~\bibnamefont {Sousa}}, \bibinfo {author} {\bibfnamefont {M.}~\bibnamefont {Copelli}},\ and\ \bibinfo {author} {\bibfnamefont {P.~V.}\ \bibnamefont {Carelli}},\ }\bibfield  {title} {\bibinfo {title} {Signatures of brain criticality unveiled by maximum entropy analysis across cortical states},\ }\href {https://doi.org/10.1103/PHYSREVE.102.012408/FIGURES/11/MEDIUM} {\bibfield  {journal} {\bibinfo  {journal} {Phys. Rev. E}\ }\textbf {\bibinfo {volume} {102}},\ \bibinfo {pages} {012408} (\bibinfo {year} {2020})}\BibitemShut {NoStop}%
\bibitem [{\citenamefont {Ioffe}\ and\ \citenamefont {Berry~II}(2017)}]{Ioffe}%
  \BibitemOpen
  \bibfield  {author} {\bibinfo {author} {\bibfnamefont {M.~L.}\ \bibnamefont {Ioffe}}\ and\ \bibinfo {author} {\bibfnamefont {M.~J.}\ \bibnamefont {Berry~II}},\ }\bibfield  {title} {\bibinfo {title} {The structured ‘low temperature’ phase of the retinal population code},\ }\href@noop {} {\bibfield  {journal} {\bibinfo  {journal} {PLoS Comput. Biol.}\ }\textbf {\bibinfo {volume} {13}},\ \bibinfo {pages} {e1005792} (\bibinfo {year} {2017})}\BibitemShut {NoStop}%
\bibitem [{\citenamefont {Morcos}\ \emph {et~al.}(2011)\citenamefont {Morcos}, \citenamefont {Pagnani}, \citenamefont {Lunt}, \citenamefont {Bertolino}, \citenamefont {Marks}, \citenamefont {Sander}, \citenamefont {Zecchina}, \citenamefont {Onuchic}, \citenamefont {Hwa},\ and\ \citenamefont {Weigt}}]{Morcos2011}%
  \BibitemOpen
  \bibfield  {author} {\bibinfo {author} {\bibfnamefont {F.}~\bibnamefont {Morcos}}, \bibinfo {author} {\bibfnamefont {A.}~\bibnamefont {Pagnani}}, \bibinfo {author} {\bibfnamefont {B.}~\bibnamefont {Lunt}}, \bibinfo {author} {\bibfnamefont {A.}~\bibnamefont {Bertolino}}, \bibinfo {author} {\bibfnamefont {D.~S.}\ \bibnamefont {Marks}}, \bibinfo {author} {\bibfnamefont {C.}~\bibnamefont {Sander}}, \bibinfo {author} {\bibfnamefont {R.}~\bibnamefont {Zecchina}}, \bibinfo {author} {\bibfnamefont {J.~N.}\ \bibnamefont {Onuchic}}, \bibinfo {author} {\bibfnamefont {T.}~\bibnamefont {Hwa}},\ and\ \bibinfo {author} {\bibfnamefont {M.}~\bibnamefont {Weigt}},\ }\bibfield  {title} {\bibinfo {title} {Direct-coupling analysis of residue coevolution captures native contacts across many protein families},\ }\href {https://doi.org/10.1073/pnas.1111471108} {\bibfield  {journal} {\bibinfo  {journal} {Proc. Natl. Acad. Sci. U. S. A.}\ }\textbf {\bibinfo {volume} {108}},\ \bibinfo {pages} {E1293} (\bibinfo {year} {2011})}\BibitemShut {NoStop}%
\bibitem [{\citenamefont {Weigt}\ \emph {et~al.}(2009)\citenamefont {Weigt}, \citenamefont {White}, \citenamefont {Szurmant}, \citenamefont {Hoch},\ and\ \citenamefont {Hwa}}]{Weigt2009}%
  \BibitemOpen
  \bibfield  {author} {\bibinfo {author} {\bibfnamefont {M.}~\bibnamefont {Weigt}}, \bibinfo {author} {\bibfnamefont {R.~A.}\ \bibnamefont {White}}, \bibinfo {author} {\bibfnamefont {H.}~\bibnamefont {Szurmant}}, \bibinfo {author} {\bibfnamefont {J.~A.}\ \bibnamefont {Hoch}},\ and\ \bibinfo {author} {\bibfnamefont {T.}~\bibnamefont {Hwa}},\ }\bibfield  {title} {\bibinfo {title} {Identification of direct residue contacts in protein-protein interaction by message passing},\ }\href {https://doi.org/10.1073/pnas.0805923106} {\bibfield  {journal} {\bibinfo  {journal} {Proc. Natl. Acad. Sci. U. S. A.}\ }\textbf {\bibinfo {volume} {106}},\ \bibinfo {pages} {67} (\bibinfo {year} {2009})}\BibitemShut {NoStop}%
\bibitem [{\citenamefont {Stein}\ \emph {et~al.}(2015)\citenamefont {Stein}, \citenamefont {Marks},\ and\ \citenamefont {Sander}}]{Stein2015}%
  \BibitemOpen
  \bibfield  {author} {\bibinfo {author} {\bibfnamefont {R.~R.}\ \bibnamefont {Stein}}, \bibinfo {author} {\bibfnamefont {D.~S.}\ \bibnamefont {Marks}},\ and\ \bibinfo {author} {\bibfnamefont {C.}~\bibnamefont {Sander}},\ }\bibfield  {title} {\bibinfo {title} {Inferring pairwise interactions from biological data using maximum-entropy probability models},\ }\href {https://doi.org/10.1371/journal.pcbi.1004182} {\bibfield  {journal} {\bibinfo  {journal} {PLoS Comput. Biol.}\ }\textbf {\bibinfo {volume} {11}},\ \bibinfo {pages} {e1004182} (\bibinfo {year} {2015})}\BibitemShut {NoStop}%
\bibitem [{\citenamefont {Lezon}\ \emph {et~al.}(2006)\citenamefont {Lezon}, \citenamefont {Banavar}, \citenamefont {Cieplak}, \citenamefont {Maritan},\ and\ \citenamefont {Fedoroff}}]{Lezon2006}%
  \BibitemOpen
  \bibfield  {author} {\bibinfo {author} {\bibfnamefont {T.~R.}\ \bibnamefont {Lezon}}, \bibinfo {author} {\bibfnamefont {J.~R.}\ \bibnamefont {Banavar}}, \bibinfo {author} {\bibfnamefont {M.}~\bibnamefont {Cieplak}}, \bibinfo {author} {\bibfnamefont {A.}~\bibnamefont {Maritan}},\ and\ \bibinfo {author} {\bibfnamefont {N.~V.}\ \bibnamefont {Fedoroff}},\ }\bibfield  {title} {\bibinfo {title} {Using the principle of entropy maximization to infer genetic interaction networks from gene expression patterns},\ }\href {https://doi.org/10.1073/PNAS.0609152103/SUPPL_FILE/09152FIG9.JPG} {\bibfield  {journal} {\bibinfo  {journal} {Proc. Natl. Acad. Sci. U. S. A.}\ }\textbf {\bibinfo {volume} {103}},\ \bibinfo {pages} {19033} (\bibinfo {year} {2006})}\BibitemShut {NoStop}%
\bibitem [{\citenamefont {Locasale}\ and\ \citenamefont {Wolf-Yadlin}(2009)}]{Locasale2009}%
  \BibitemOpen
  \bibfield  {author} {\bibinfo {author} {\bibfnamefont {J.~W.}\ \bibnamefont {Locasale}}\ and\ \bibinfo {author} {\bibfnamefont {A.}~\bibnamefont {Wolf-Yadlin}},\ }\bibfield  {title} {\bibinfo {title} {Maximum entropy reconstructions of dynamic signaling networks from quantitative proteomics data},\ }\href {https://doi.org/10.1371/journal.pone.0006522} {\bibfield  {journal} {\bibinfo  {journal} {PLoS ONE}\ }\textbf {\bibinfo {volume} {4}},\ \bibinfo {pages} {e6522} (\bibinfo {year} {2009})}\BibitemShut {NoStop}%
\bibitem [{\citenamefont {Bialek}\ \emph {et~al.}(2012)\citenamefont {Bialek}, \citenamefont {Cavagna}, \citenamefont {Giardina}, \citenamefont {Mora}, \citenamefont {Silvestri}, \citenamefont {Viale},\ and\ \citenamefont {Walczak}}]{Bialek2012}%
  \BibitemOpen
  \bibfield  {author} {\bibinfo {author} {\bibfnamefont {W.}~\bibnamefont {Bialek}}, \bibinfo {author} {\bibfnamefont {A.}~\bibnamefont {Cavagna}}, \bibinfo {author} {\bibfnamefont {I.}~\bibnamefont {Giardina}}, \bibinfo {author} {\bibfnamefont {T.}~\bibnamefont {Mora}}, \bibinfo {author} {\bibfnamefont {E.}~\bibnamefont {Silvestri}}, \bibinfo {author} {\bibfnamefont {M.}~\bibnamefont {Viale}},\ and\ \bibinfo {author} {\bibfnamefont {A.~M.}\ \bibnamefont {Walczak}},\ }\bibfield  {title} {\bibinfo {title} {Statistical mechanics for natural flocks of birds},\ }\href {https://doi.org/10.1073/pnas.1118633109} {\bibfield  {journal} {\bibinfo  {journal} {Proc. Natl. Acad. Sci. U. S. A.}\ }\textbf {\bibinfo {volume} {109}},\ \bibinfo {pages} {4786} (\bibinfo {year} {2012})}\BibitemShut {NoStop}%
\bibitem [{\citenamefont {Bialek}\ \emph {et~al.}(2014)\citenamefont {Bialek}, \citenamefont {Cavagna}, \citenamefont {Giardina}, \citenamefont {Mora}, \citenamefont {Pohl}, \citenamefont {Silvestri}, \citenamefont {Viale},\ and\ \citenamefont {Walczak}}]{Bialek2014}%
  \BibitemOpen
  \bibfield  {author} {\bibinfo {author} {\bibfnamefont {W.}~\bibnamefont {Bialek}}, \bibinfo {author} {\bibfnamefont {A.}~\bibnamefont {Cavagna}}, \bibinfo {author} {\bibfnamefont {I.}~\bibnamefont {Giardina}}, \bibinfo {author} {\bibfnamefont {T.}~\bibnamefont {Mora}}, \bibinfo {author} {\bibfnamefont {O.}~\bibnamefont {Pohl}}, \bibinfo {author} {\bibfnamefont {E.}~\bibnamefont {Silvestri}}, \bibinfo {author} {\bibfnamefont {M.}~\bibnamefont {Viale}},\ and\ \bibinfo {author} {\bibfnamefont {A.~M.}\ \bibnamefont {Walczak}},\ }\bibfield  {title} {\bibinfo {title} {Social interactions dominate speed control in poising natural flocks near criticality},\ }\href {https://doi.org/10.1073/PNAS.1324045111/SUPPL_FILE/PNAS.201324045SI.PDF} {\bibfield  {journal} {\bibinfo  {journal} {Proc. Natl. Acad. Sci. U. S. A.}\ }\textbf {\bibinfo {volume} {111}},\ \bibinfo {pages} {7212} (\bibinfo {year} {2014})}\BibitemShut {NoStop}%
\bibitem [{\citenamefont {Burleson-Lesser}\ \emph {et~al.}(2017)\citenamefont {Burleson-Lesser}, \citenamefont {Morone}, \citenamefont {DeGuzman}, \citenamefont {Parra},\ and\ \citenamefont {Makse}}]{Burleson2017}%
  \BibitemOpen
  \bibfield  {author} {\bibinfo {author} {\bibfnamefont {K.}~\bibnamefont {Burleson-Lesser}}, \bibinfo {author} {\bibfnamefont {F.}~\bibnamefont {Morone}}, \bibinfo {author} {\bibfnamefont {P.}~\bibnamefont {DeGuzman}}, \bibinfo {author} {\bibfnamefont {L.~C.}\ \bibnamefont {Parra}},\ and\ \bibinfo {author} {\bibfnamefont {H.~A.}\ \bibnamefont {Makse}},\ }\bibfield  {title} {\bibinfo {title} {Collective behaviour in video viewing: A thermodynamic analysis of gaze position},\ }\href {https://doi.org/10.1371/journal.pone.0168995} {\bibfield  {journal} {\bibinfo  {journal} {PLoS ONE}\ }\textbf {\bibinfo {volume} {12}},\ \bibinfo {pages} {1} (\bibinfo {year} {2017})}\BibitemShut {NoStop}%
\bibitem [{\citenamefont {Torres}\ \emph {et~al.}(2021)\citenamefont {Torres}, \citenamefont {Sena}, \citenamefont {Carmona}, \citenamefont {Moreira}, \citenamefont {Makse},\ and\ \citenamefont {Andrade}}]{Debora2021}%
  \BibitemOpen
  \bibfield  {author} {\bibinfo {author} {\bibfnamefont {D.}~\bibnamefont {Torres}}, \bibinfo {author} {\bibfnamefont {W.~R.}\ \bibnamefont {Sena}}, \bibinfo {author} {\bibfnamefont {H.~A.}\ \bibnamefont {Carmona}}, \bibinfo {author} {\bibfnamefont {A.~A.}\ \bibnamefont {Moreira}}, \bibinfo {author} {\bibfnamefont {H.~A.}\ \bibnamefont {Makse}},\ and\ \bibinfo {author} {\bibfnamefont {J.~S.}\ \bibnamefont {Andrade}},\ }\bibfield  {title} {\bibinfo {title} {Eye-tracking as a proxy for coherence and complexity of texts},\ }\href {https://doi.org/10.1371/journal.pone.0260236} {\bibfield  {journal} {\bibinfo  {journal} {PLoS ONE}\ }\textbf {\bibinfo {volume} {16}},\ \bibinfo {pages} {e0260236} (\bibinfo {year} {2021})}\BibitemShut {NoStop}%
\bibitem [{\citenamefont {Bury}(2013)}]{Bury2013}%
  \BibitemOpen
  \bibfield  {author} {\bibinfo {author} {\bibfnamefont {T.}~\bibnamefont {Bury}},\ }\bibfield  {title} {\bibinfo {title} {A statistical physics perspective on criticality in financial markets},\ }\href {https://doi.org/10.1088/1742-5468/2013/11/P11004} {\bibfield  {journal} {\bibinfo  {journal} {J. Stat. Mech.}\ }\textbf {\bibinfo {volume} {2013}},\ \bibinfo {pages} {P11004} (\bibinfo {year} {2013})}\BibitemShut {NoStop}%
\bibitem [{\citenamefont {Mora}\ and\ \citenamefont {Bialek}(2011)}]{Mora2011}%
  \BibitemOpen
  \bibfield  {author} {\bibinfo {author} {\bibfnamefont {T.}~\bibnamefont {Mora}}\ and\ \bibinfo {author} {\bibfnamefont {W.}~\bibnamefont {Bialek}},\ }\bibfield  {title} {\bibinfo {title} {Are {Biological} {Systems} {Poised} at {Criticality}?},\ }\href {https://doi.org/10.1007/s10955-011-0229-4} {\bibfield  {journal} {\bibinfo  {journal} {J. Stat. Phys.}\ }\textbf {\bibinfo {volume} {144}},\ \bibinfo {pages} {268} (\bibinfo {year} {2011})}\BibitemShut {NoStop}%
\bibitem [{\citenamefont {da~Fazenda}(2003)}]{nfe2023}%
  \BibitemOpen
  \bibfield  {author} {\bibinfo {author} {\bibfnamefont {M.}~\bibnamefont {da~Fazenda}},\ }\href {https://www.nfe.fazenda.gov.br/portal/principal.aspx} {\bibinfo {title} {Portal da nota fiscal eletrônica}} (\bibinfo {year} {2003}),\ \bibinfo {note} {accessed 2023-05-12}\BibitemShut {NoStop}%
\bibitem [{\citenamefont {Fischler}\ and\ \citenamefont {Bolles}(1981)}]{Fischler1981}%
  \BibitemOpen
  \bibfield  {author} {\bibinfo {author} {\bibfnamefont {M.~A.}\ \bibnamefont {Fischler}}\ and\ \bibinfo {author} {\bibfnamefont {R.~C.}\ \bibnamefont {Bolles}},\ }\bibfield  {title} {\bibinfo {title} {Random sample consensus},\ }\href {https://doi.org/10.1145/358669.358692} {\bibfield  {journal} {\bibinfo  {journal} {Commun. ACM}\ }\textbf {\bibinfo {volume} {24}},\ \bibinfo {pages} {381} (\bibinfo {year} {1981})}\BibitemShut {NoStop}%
\bibitem [{\citenamefont {Chum}\ and\ \citenamefont {Matas}(2008)}]{Chum2008}%
  \BibitemOpen
  \bibfield  {author} {\bibinfo {author} {\bibfnamefont {O.}~\bibnamefont {Chum}}\ and\ \bibinfo {author} {\bibfnamefont {J.}~\bibnamefont {Matas}},\ }\bibfield  {title} {\bibinfo {title} {Optimal randomized ransac},\ }\href {https://doi.org/10.1109/TPAMI.2007.70787} {\bibfield  {journal} {\bibinfo  {journal} {IEEE Trans. Pattern Anal. Mach. Intell.}\ }\textbf {\bibinfo {volume} {30}},\ \bibinfo {pages} {1472} (\bibinfo {year} {2008})}\BibitemShut {NoStop}%
\bibitem [{\citenamefont {Newman}(2004)}]{Newman2004}%
  \BibitemOpen
  \bibfield  {author} {\bibinfo {author} {\bibfnamefont {M.~E.}\ \bibnamefont {Newman}},\ }\bibfield  {title} {\bibinfo {title} {Analysis of weighted networks},\ }\href {https://doi.org/10.1103/PHYSREVE.70.056131/FIGURES/3/MEDIUM} {\bibfield  {journal} {\bibinfo  {journal} {Phys. Rev. E}\ }\textbf {\bibinfo {volume} {70}},\ \bibinfo {pages} {9} (\bibinfo {year} {2004})}\BibitemShut {NoStop}%
\bibitem [{\citenamefont {Barrat}\ \emph {et~al.}(2004)\citenamefont {Barrat}, \citenamefont {Barthélemy}, \citenamefont {Pastor-Satorras},\ and\ \citenamefont {Vespignani}}]{Barrat2004}%
  \BibitemOpen
  \bibfield  {author} {\bibinfo {author} {\bibfnamefont {A.}~\bibnamefont {Barrat}}, \bibinfo {author} {\bibfnamefont {M.}~\bibnamefont {Barthélemy}}, \bibinfo {author} {\bibfnamefont {R.}~\bibnamefont {Pastor-Satorras}},\ and\ \bibinfo {author} {\bibfnamefont {A.}~\bibnamefont {Vespignani}},\ }\bibfield  {title} {\bibinfo {title} {The architecture of complex weighted networks},\ }\href {https://doi.org/10.1073/PNAS.0400087101/ASSET/7970A6FD-0C68-49D2-AB37-938124DCEAFA/ASSETS/GRAPHIC/ZPQ0080439150007.JPEG} {\bibfield  {journal} {\bibinfo  {journal} {Proc. Natl. Acad. Sci. U. S. A.}\ }\textbf {\bibinfo {volume} {101}},\ \bibinfo {pages} {3747} (\bibinfo {year} {2004})}\BibitemShut {NoStop}%
\bibitem [{\citenamefont {Rubinov}\ and\ \citenamefont {Sporns}(2010)}]{Rubinov2010}%
  \BibitemOpen
  \bibfield  {author} {\bibinfo {author} {\bibfnamefont {M.}~\bibnamefont {Rubinov}}\ and\ \bibinfo {author} {\bibfnamefont {O.}~\bibnamefont {Sporns}},\ }\bibfield  {title} {\bibinfo {title} {Complex network measures of brain connectivity: Uses and interpretations},\ }\href {https://doi.org/10.1016/J.NEUROIMAGE.2009.10.003} {\bibfield  {journal} {\bibinfo  {journal} {NeuroImage}\ }\textbf {\bibinfo {volume} {52}},\ \bibinfo {pages} {1059} (\bibinfo {year} {2010})}\BibitemShut {NoStop}%
\bibitem [{\citenamefont {Boccaletti}\ \emph {et~al.}(2014)\citenamefont {Boccaletti}, \citenamefont {Bianconi}, \citenamefont {Criado}, \citenamefont {del Genio}, \citenamefont {Gómez-Gardeñes}, \citenamefont {Romance}, \citenamefont {Sendiña-Nadal}, \citenamefont {Wang},\ and\ \citenamefont {Zanin}}]{Boccaletti2014}%
  \BibitemOpen
  \bibfield  {author} {\bibinfo {author} {\bibfnamefont {S.}~\bibnamefont {Boccaletti}}, \bibinfo {author} {\bibfnamefont {G.}~\bibnamefont {Bianconi}}, \bibinfo {author} {\bibfnamefont {R.}~\bibnamefont {Criado}}, \bibinfo {author} {\bibfnamefont {C.~I.}\ \bibnamefont {del Genio}}, \bibinfo {author} {\bibfnamefont {J.}~\bibnamefont {Gómez-Gardeñes}}, \bibinfo {author} {\bibfnamefont {M.}~\bibnamefont {Romance}}, \bibinfo {author} {\bibfnamefont {I.}~\bibnamefont {Sendiña-Nadal}}, \bibinfo {author} {\bibfnamefont {Z.}~\bibnamefont {Wang}},\ and\ \bibinfo {author} {\bibfnamefont {M.}~\bibnamefont {Zanin}},\ }\bibfield  {title} {\bibinfo {title} {The structure and dynamics of multilayer networks},\ }\href {https://doi.org/10.1016/J.PHYSREP.2014.07.001} {\bibfield  {journal} {\bibinfo  {journal} {Phys. Rep.}\ }\textbf {\bibinfo {volume} {544}},\ \bibinfo {pages} {1} (\bibinfo {year} {2014})}\BibitemShut {NoStop}%
\bibitem [{\citenamefont {Rosvall}\ and\ \citenamefont {Bergstrom}(2008)}]{Rosvall2008}%
  \BibitemOpen
  \bibfield  {author} {\bibinfo {author} {\bibfnamefont {M.}~\bibnamefont {Rosvall}}\ and\ \bibinfo {author} {\bibfnamefont {C.~T.}\ \bibnamefont {Bergstrom}},\ }\bibfield  {title} {\bibinfo {title} {Maps of random walks on complex networks reveal community structure},\ }\href {https://doi.org/10.1073/pnas.0706851105} {\bibfield  {journal} {\bibinfo  {journal} {Proc. Natl. Acad. Sci. U. S. A.}\ }\textbf {\bibinfo {volume} {105}},\ \bibinfo {pages} {1118} (\bibinfo {year} {2008})}\BibitemShut {NoStop}%
\bibitem [{\citenamefont {Rosvall}\ and\ \citenamefont {Bergstrom}(2011)}]{Rosvall2011}%
  \BibitemOpen
  \bibfield  {author} {\bibinfo {author} {\bibfnamefont {M.}~\bibnamefont {Rosvall}}\ and\ \bibinfo {author} {\bibfnamefont {C.~T.}\ \bibnamefont {Bergstrom}},\ }\bibfield  {title} {\bibinfo {title} {Multilevel compression of random walks on networks reveals hierarchical organization in large integrated systems},\ }\href {https://doi.org/10.1371/JOURNAL.PONE.0018209} {\bibfield  {journal} {\bibinfo  {journal} {PLoS ONE}\ }\textbf {\bibinfo {volume} {6}},\ \bibinfo {pages} {e18209} (\bibinfo {year} {2011})}\BibitemShut {NoStop}%
\bibitem [{\citenamefont {Alzahrani}\ and\ \citenamefont {Horadam}(2016)}]{Alzahrani2016}%
  \BibitemOpen
  \bibfield  {author} {\bibinfo {author} {\bibfnamefont {T.}~\bibnamefont {Alzahrani}}\ and\ \bibinfo {author} {\bibfnamefont {K.~J.}\ \bibnamefont {Horadam}},\ }\bibfield  {title} {\bibinfo {title} {Community detection in bipartite networks: Algorithms and case studies},\ }in\ \href {https://doi.org/10.1007/978-3-662-47824-0_2} {\emph {\bibinfo {booktitle} {Complex Systems and Networks: Dynamics, Controls and Applications}}},\ \bibinfo {editor} {edited by\ \bibinfo {editor} {\bibfnamefont {J.}~\bibnamefont {Lü}}, \bibinfo {editor} {\bibfnamefont {X.}~\bibnamefont {Yu}}, \bibinfo {editor} {\bibfnamefont {G.}~\bibnamefont {Chen}},\ and\ \bibinfo {editor} {\bibfnamefont {W.}~\bibnamefont {Yu}}}\ (\bibinfo  {publisher} {Springer Berlin Heidelberg},\ \bibinfo {year} {2016})\ pp.\ \bibinfo {pages} {25--50}\BibitemShut {NoStop}%
\bibitem [{\citenamefont {Nasser}\ and\ \citenamefont {Vuorinen}(2020)}]{Nasser2020}%
  \BibitemOpen
  \bibfield  {author} {\bibinfo {author} {\bibfnamefont {M.~M.~S.}\ \bibnamefont {Nasser}}\ and\ \bibinfo {author} {\bibfnamefont {M.}~\bibnamefont {Vuorinen}},\ }\bibfield  {title} {\bibinfo {title} {Conformal invariants in simply connected domains},\ }\href {https://doi.org/10.1007/s40315-020-00351-8} {\bibfield  {journal} {\bibinfo  {journal} {Lect. Notes. Math.}\ }\textbf {\bibinfo {volume} {20}},\ \bibinfo {pages} {747} (\bibinfo {year} {2020})}\BibitemShut {NoStop}%
\bibitem [{\citenamefont {Papamichael}\ and\ \citenamefont {Kokkinos}(1981)}]{Papamichael1981}%
  \BibitemOpen
  \bibfield  {author} {\bibinfo {author} {\bibfnamefont {N.}~\bibnamefont {Papamichael}}\ and\ \bibinfo {author} {\bibfnamefont {C.~A.}\ \bibnamefont {Kokkinos}},\ }\bibfield  {title} {\bibinfo {title} {Two numerical methods for the conformal mapping of simply-connected domains},\ }\href@noop {} {\bibfield  {journal} {\bibinfo  {journal} {Comput. Methods Appl. Mech. Eng.}\ }\textbf {\bibinfo {volume} {28}},\ \bibinfo {pages} {285} (\bibinfo {year} {1981})}\BibitemShut {NoStop}%
\bibitem [{\citenamefont {Fortunato}\ and\ \citenamefont {Hric}(2016)}]{Fortunato2016}%
  \BibitemOpen
  \bibfield  {author} {\bibinfo {author} {\bibfnamefont {S.}~\bibnamefont {Fortunato}}\ and\ \bibinfo {author} {\bibfnamefont {D.}~\bibnamefont {Hric}},\ }\bibfield  {title} {\bibinfo {title} {Community detection in networks: A user guide},\ }\href {https://doi.org/10.1016/J.PHYSREP.2016.09.002} {\bibfield  {journal} {\bibinfo  {journal} {Phys. Rep.}\ }\textbf {\bibinfo {volume} {659}},\ \bibinfo {pages} {1} (\bibinfo {year} {2016})}\BibitemShut {NoStop}%
\bibitem [{\citenamefont {Bernard}\ \emph {et~al.}(2007)\citenamefont {Bernard}, \citenamefont {Redding},\ and\ \citenamefont {Schott}}]{Bernard2007}%
  \BibitemOpen
  \bibfield  {author} {\bibinfo {author} {\bibfnamefont {A.~B.}\ \bibnamefont {Bernard}}, \bibinfo {author} {\bibfnamefont {S.~J.}\ \bibnamefont {Redding}},\ and\ \bibinfo {author} {\bibfnamefont {P.~K.}\ \bibnamefont {Schott}},\ }\bibfield  {title} {\bibinfo {title} {Comparative advantage and heterogeneous firms},\ }\href {https://doi.org/10.1111/J.1467-937X.2007.00413.X} {\bibfield  {journal} {\bibinfo  {journal} {Rev. Econ. Stud.}\ }\textbf {\bibinfo {volume} {74}},\ \bibinfo {pages} {31} (\bibinfo {year} {2007})}\BibitemShut {NoStop}%
\bibitem [{\citenamefont {Schneidman}\ \emph {et~al.}(2006)\citenamefont {Schneidman}, \citenamefont {Berry}, \citenamefont {Segev},\ and\ \citenamefont {Bialek}}]{Schneidman2006}%
  \BibitemOpen
  \bibfield  {author} {\bibinfo {author} {\bibfnamefont {E.}~\bibnamefont {Schneidman}}, \bibinfo {author} {\bibfnamefont {M.~J.}\ \bibnamefont {Berry}}, \bibinfo {author} {\bibfnamefont {R.}~\bibnamefont {Segev}},\ and\ \bibinfo {author} {\bibfnamefont {W.}~\bibnamefont {Bialek}},\ }\bibfield  {title} {\bibinfo {title} {Weak pairwise correlations imply strongly correlated network states in a neural population},\ }\href {https://doi.org/10.1038/nature04701} {\bibfield  {journal} {\bibinfo  {journal} {Nature}\ }\textbf {\bibinfo {volume} {440}},\ \bibinfo {pages} {1007} (\bibinfo {year} {2006})}\BibitemShut {NoStop}%
\bibitem [{\citenamefont {Tkačik}\ \emph {et~al.}(2014)\citenamefont {Tkačik}, \citenamefont {Marre}, \citenamefont {Amodei}, \citenamefont {Schneidman}, \citenamefont {Bialek},\ and\ \citenamefont {Berry}}]{Tkacik2014}%
  \BibitemOpen
  \bibfield  {author} {\bibinfo {author} {\bibfnamefont {G.}~\bibnamefont {Tkačik}}, \bibinfo {author} {\bibfnamefont {O.}~\bibnamefont {Marre}}, \bibinfo {author} {\bibfnamefont {D.}~\bibnamefont {Amodei}}, \bibinfo {author} {\bibfnamefont {E.}~\bibnamefont {Schneidman}}, \bibinfo {author} {\bibfnamefont {W.}~\bibnamefont {Bialek}},\ and\ \bibinfo {author} {\bibfnamefont {M.~J.}\ \bibnamefont {Berry}},\ }\bibfield  {title} {\bibinfo {title} {Searching for collective behavior in a large network of sensory neurons},\ }\href {https://doi.org/10.1371/JOURNAL.PCBI.1003408} {\bibfield  {journal} {\bibinfo  {journal} {PLoS Comput. Biol.}\ }\textbf {\bibinfo {volume} {10}},\ \bibinfo {pages} {e1003408} (\bibinfo {year} {2014})}\BibitemShut {NoStop}%
\bibitem [{\citenamefont {Tkačik}\ \emph {et~al.}(2015)\citenamefont {Tkačik}, \citenamefont {Mora}, \citenamefont {Marre}, \citenamefont {Amodei}, \citenamefont {Palmer}, \citenamefont {Berry},\ and\ \citenamefont {Bialek}}]{Tkacik2015}%
  \BibitemOpen
  \bibfield  {author} {\bibinfo {author} {\bibfnamefont {G.}~\bibnamefont {Tkačik}}, \bibinfo {author} {\bibfnamefont {T.}~\bibnamefont {Mora}}, \bibinfo {author} {\bibfnamefont {O.}~\bibnamefont {Marre}}, \bibinfo {author} {\bibfnamefont {D.}~\bibnamefont {Amodei}}, \bibinfo {author} {\bibfnamefont {S.~E.}\ \bibnamefont {Palmer}}, \bibinfo {author} {\bibfnamefont {M.~J.}\ \bibnamefont {Berry}},\ and\ \bibinfo {author} {\bibfnamefont {W.}~\bibnamefont {Bialek}},\ }\bibfield  {title} {\bibinfo {title} {Thermodynamics and signatures of criticality in a network of neurons},\ }\href {https://doi.org/10.1073/PNAS.1514188112/SUPPL_FILE/PNAS.201514188SI.PDF} {\bibfield  {journal} {\bibinfo  {journal} {Proc. Natl. Acad. Sci. U. S. A.}\ }\textbf {\bibinfo {volume} {112}},\ \bibinfo {pages} {11508} (\bibinfo {year} {2015})}\BibitemShut {NoStop}%
\bibitem [{\citenamefont {Bialek}\ \emph {et~al.}(2013)\citenamefont {Bialek}, \citenamefont {Cavagna}, \citenamefont {Giardina}, \citenamefont {Mora}, \citenamefont {Pohl}, \citenamefont {Silvestri}, \citenamefont {Viale},\ and\ \citenamefont {Walczak}}]{Bialek2013}%
  \BibitemOpen
  \bibfield  {author} {\bibinfo {author} {\bibfnamefont {W.}~\bibnamefont {Bialek}}, \bibinfo {author} {\bibfnamefont {A.}~\bibnamefont {Cavagna}}, \bibinfo {author} {\bibfnamefont {I.}~\bibnamefont {Giardina}}, \bibinfo {author} {\bibfnamefont {T.}~\bibnamefont {Mora}}, \bibinfo {author} {\bibfnamefont {O.}~\bibnamefont {Pohl}}, \bibinfo {author} {\bibfnamefont {E.}~\bibnamefont {Silvestri}}, \bibinfo {author} {\bibfnamefont {M.}~\bibnamefont {Viale}},\ and\ \bibinfo {author} {\bibfnamefont {A.}~\bibnamefont {Walczak}},\ }\bibfield  {title} {\bibinfo {title} {Social interactions dominate speed control in driving natural flocks toward criticality},\ }\href {https://doi.org/10.1073/pnas.1324045111} {\bibfield  {journal} {\bibinfo  {journal} {Proc. Natl. Acad. Sci. U. S. A.}\ }\textbf {\bibinfo {volume} {111}},\ \bibinfo {pages} {7212} (\bibinfo {year} {2013})}\BibitemShut {NoStop}%
\bibitem [{\citenamefont {Nguyen}\ \emph {et~al.}(2017)\citenamefont {Nguyen}, \citenamefont {Zecchina},\ and\ \citenamefont {Berg}}]{Nguyen2017}%
  \BibitemOpen
  \bibfield  {author} {\bibinfo {author} {\bibfnamefont {H.~C.}\ \bibnamefont {Nguyen}}, \bibinfo {author} {\bibfnamefont {R.}~\bibnamefont {Zecchina}},\ and\ \bibinfo {author} {\bibfnamefont {J.}~\bibnamefont {Berg}},\ }\bibfield  {title} {\bibinfo {title} {Inverse statistical problems: from the inverse ising problem to data science},\ }\href {https://doi.org/10.1080/00018732.2017.1341604} {\bibfield  {journal} {\bibinfo  {journal} {Adv. Phys.}\ }\textbf {\bibinfo {volume} {66}},\ \bibinfo {pages} {197} (\bibinfo {year} {2017})}\BibitemShut {NoStop}%
\bibitem [{\citenamefont {Ataliba}\ \emph {et~al.}(2014)\citenamefont {Ataliba}, \citenamefont {Barreto}, \citenamefont {Sarquis},\ and\ \citenamefont {Menezes}}]{Ataliba2014}%
  \BibitemOpen
  \bibfield  {author} {\bibinfo {author} {\bibfnamefont {F.}~\bibnamefont {Ataliba}}, \bibinfo {author} {\bibfnamefont {F.~D.}\ \bibnamefont {Barreto}}, \bibinfo {author} {\bibfnamefont {A.}~\bibnamefont {Sarquis}},\ and\ \bibinfo {author} {\bibfnamefont {B.~D.}\ \bibnamefont {Menezes}},\ }\href {www.ipece.ce.gov.br} {\bibinfo {title} {Desenvolvimento econômico do {Ceará}: Evidências recentes e reflexões}} (\bibinfo {year} {2014})\BibitemShut {NoStop}%
\end{thebibliography}
\end{document}